\documentclass[review]{elsarticle}

\usepackage{bm}
\usepackage{amsmath, wasysym}
\usepackage{amsfonts}%
\usepackage{amssymb}%
\usepackage{color}
\usepackage{hyperref}
\usepackage{verbatim} 
\usepackage{epsfig,epsf}
\usepackage{graphicx,tabularx}
\newcolumntype{Y}{>{\centering\arraybackslash}X}
\usepackage{epstopdf}
\usepackage{lineno}
\usepackage{mathrsfs}
\usepackage{ulem}
\usepackage[percent]{overpic}
\modulolinenumbers[5]

\journal{....}

\begin{document}

\begin{frontmatter}

\title{In silico evidence that protein unfolding is as a precursor of the protein aggregation}

\author{Valentino Bianco}
\address{Faculty of Chemistry, Chemical Physics Deprtment, Universidad Complutense de Madrid, Plaza de las Ciencias, Ciudad Universitaria, Madrid 28040, Spain}
\ead{vabianco@ucm.es}
\author{Giancarlo Franzese}
\address{Secci\'o de
 F\'isica Estad\'istica i Interdisciplin\`aria--Departament de F\'isica
 de la Mat\`eria Condensada, Facultat de F\'isica \& Institute of Nanoscience and Nanotechnology (IN2UB),
  Universitat de Barcelona, Mart\'i i Franqu\`es 1, 08028 Barcelona, Spain}
  \ead{gfranzese@ub.edu}
\author{Ivan Coluzza}
\address{CIC biomaGUNE, Paseo Miramon 182, 20014 San Sebastian, Spain.
IKERBASQUE, Basque Foundation for Science, 48013 Bilbao, Spain.}
\ead{icoluzza@cicbiomagune.es}



\begin{abstract}
We present a computational study on the folding and aggregation of proteins in aqueous environment, as function of its concentration.  We show how the increase of the concentration of individual protein species can induce a partial unfolding of the native conformation without the occurrence of aggregates. A further increment of the protein concentration results in the complete loss of the folded structures and induces the formation of protein aggregates. We discuss the effect of the protein interface on the water fluctuations in the protein hydration shell and their relevance in the protein-protein interaction.
\end{abstract}  

\end{frontmatter}

\section{Introduction}
Proteins cover a range of fundamental functions in the human body: i) the enzymes and hormones are proteins; ii) proteins can carry other biomolecules within the cellular environment; iii) proteins are a source of energy; iv) proteins are necessary to build and repair tissues ~\cite{Finkelstein2016}.
A protein is synthesized in the ribosome and, despite the fact that the cellular environment is very crowded, it is capable to reach its native conformation (mostly dictated by the protein sequence). This process is usually spontaneous--at least for small protein--or is driven by complex interactions with other biomolecules, like the chaperones.
Proteins can aggregate after they folded in the native state --- through the formation of chemical bonds or self-assembling --- or via  unfolded intermediate conformations and their propensity to aggregate is related to a series of factors, like the flexibility of the protein structure \cite{DeSimone2012a} or the sub-cellular volume where the protein resides \cite{Tartaglia2009}.
In particular, non-native protein aggregates are commonly formed through a multi-step process and are composed by native-like--partially folded intermediate structures   \cite{Eliezer1993, Fink1998, Roberts2007, Neudecker2012}. Inappropriate protein aggregation represents a crucial issue in biology and medicine, being associated to a growing number of diseases such as Alzheimer's and Parkinson's disease \cite{Ross2005, Chiti2006, Aguzzi2010, Knowles2014}. 
In order to guarantee the correct biological functions, proteins have evolved to have a low enough propensity to aggregate within a range of protein expression required for their biological activity, but with no margin to respond to external factors increasing/decreasing their expression/solubility \cite{Schroder2002, Tartaglia2007, Tartaglia2009}. Indeed, protein aggregation is mostly unavoidable when proteins are expressed at concentrations higher than the natural ones.

The mechanisms leading to the failure of the folding process and to the formation of potentially dangerous protein aggregates are matter of large scientific debate \cite{DOBSON20043}, where computational tools have largely contributed to elucidate some crucial aspects. Nevertheless, to date an extensive computational study of protein aggregation with all-atom simulations including the solvent explicitly remains not practicable, making the coarse-grain approach a valid tool to rationalize those complex systems \cite{Cellmer2007, Nasica-Labouze2015}. In particular, lattice models have been largely exploited to address fundamental questions on protein folding and aggregation \cite{Broglia1998, Toma2000, Bratko2001, Combe2003, Dima2002, Oakley2005, Ji2005, Cellmer2005, Zhang2008, Abeln2014a, Morriss-Andrews2015}. 
According to these studies, the presence of more than one chain leads to aggregate---although each protein contains a considerable fraction of native structure---with consequent loss of the funnel-like free-energy landscape \cite{Broglia1998, Bratko2001, Cellmer2005}.

All these studies, usually performed with a fixed sequence \cite{Broglia1998, Cellmer2007} of with Go-like models \cite{Bratko2001,Combe2003}, miss the explicit contribution of water, which instead is supposed to play an important role in the  protein-protein recognition and aggregation \cite{Chi2003, DeSimone2005, Krone2008, Thirumalai2012, Fichou2015, Arya2016}. 
Moreover, works implicitly accounting for water show that proteins with hydrophobic amino acids on the surface are prone to aggregate \cite{Zhang2008}, although in nature many proteins present a considerable fraction of hydrophobic and non-polar amino acids on their native surface.

Here we present a computational study on the folding, stability and aggregation of proteins optimized according to the environment. We consider a series of native protein structures and for each we determine one or more sequences designed to make the protein fold into the aqueous environment \cite{Bianco2017}. Each sequence exhibits a different ratio between the number of hydrophilic amino acids exposed to the solvent and the number of hydrophobic amino acids buried into the core of the protein in its native conformation.
For each protein, we study its capability to fold as function of its concentration. We show that the propensity to aggregate is not strictly related to the hydrophobicity of the protein surface.

\section{Simulation scheme}

To perform this study we adopt a coarse-grained lattice representation of proteins which is computationally affordable and has been largely adopted in leterature \cite{caldarelli2001, marques2003,  Patel2007, matysiak2012, Bianco2015, VanDijk2016a, Bianco2017}. 
A protein is represented as a self-avoiding heteropolymer, composed by 20 amino acids, interacting each other through a  
nearest-neighbour potential given by the Miyazawa Jernigan interaction matrix \cite{Miyazawa1985, Coluzza2004, Kyte1982}\footnote{The matrix has been scaled of a factor 2, increasing the effective amino acid-amino acid interaction, to account for a lower surface-volume ratio in two dimensions.}. 

The protein is embedded in water, explicitly modeled via the {\it many-body} water model which has been proven to reproduce, at least qualitatively, the thermodynamic and dynamic behavior of water \cite{Stokely2010, Mazza2011, delosSantos2011, Franzese2013, Bianco2014}, including its interplay with proteins \cite{Franzese2011, Bianco2012a, Bianco2015, Bianco2017a, Bianco2017}.
The coarse-grain representation of the water molecules, adopted to describe water at constant number of molecules $N$, constant temperature $T$ and constant pressure $P$, replaces the coordinates and orientations of the water molecules 
by a continuous density field and discrete bonding variables, respectively. The discrete variables describe the local hydrogen-bond (HB) formation and its cooperativity, leading to a local open--tetrahedral structure of the water molecules.

Since the protein is composed by hydrophilic $\zeta$ and hydrophobic $\Phi$ amino acids, we assume that the first interact with water decreasing the local energy, while the second affect the water--water HB in the $\Phi$ hydration shell
\footnote{The hydration shell is defined by the water molecules which are first-neighbors of the amino acids.}. In particular, we assume that i) the water--water HB at the $\Phi$ interface are stronger than HB formed in the bulk consistent with the observation that water-water HBs in the $\Phi$ hydration shell are more stable and more correlated with respect to the bulk HBs \cite{Dias2008,   Petersen2009, SarupriaPRL2009, Tarasevich2011, DavisNat2012, laage2017}; ii) the local density fluctuations at the $\Phi$ interface are reduced upon pressurization, as observed in \cite{Sarupria2009, Das2012, Ghosh2001, Dias2014}.

A detailed description of the model is reported in the nexs section, and in Ref. \cite{Bianco2015, Bianco2017a, Bianco2017}.

\begin{figure}
\centering
Protein $A_0$ \hspace{1 cm} Protein $A_1$ \hspace{1 cm} Protein $A_2$\\ 
 \includegraphics[scale=0.07]{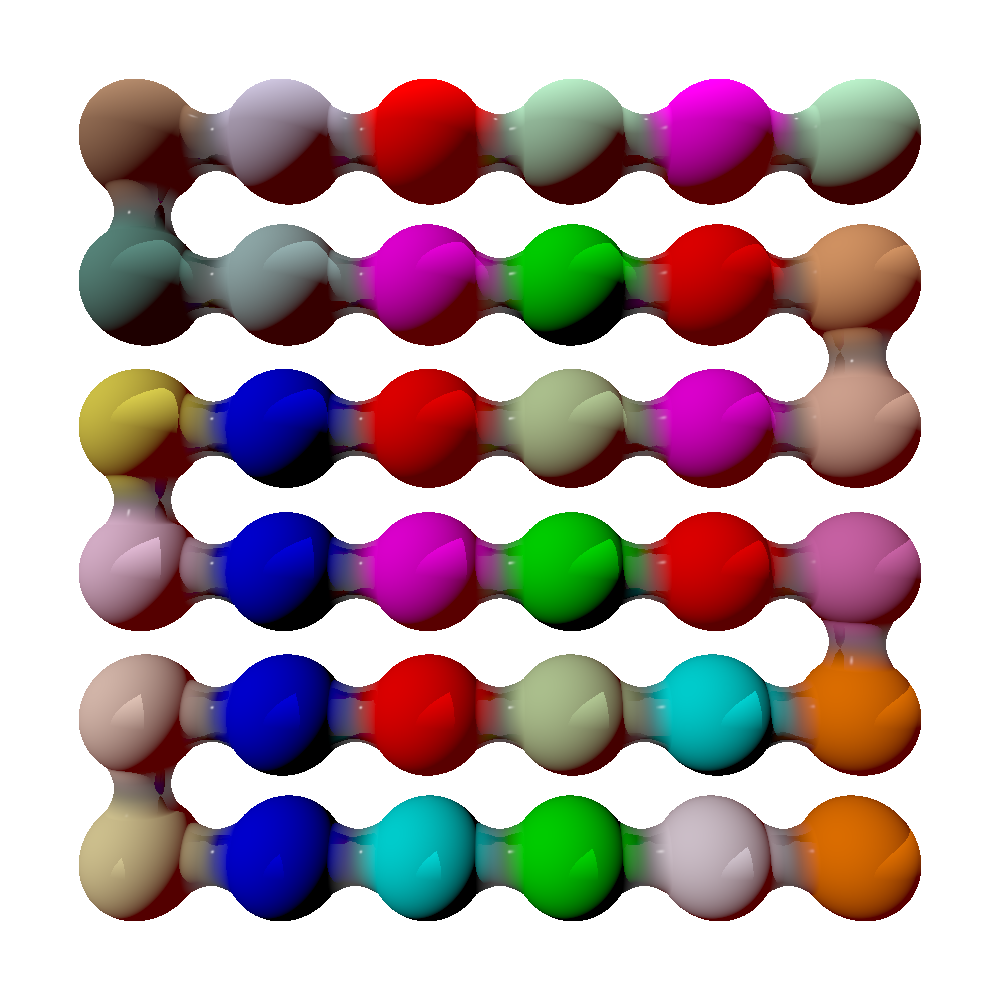}
 \includegraphics[scale=0.07]{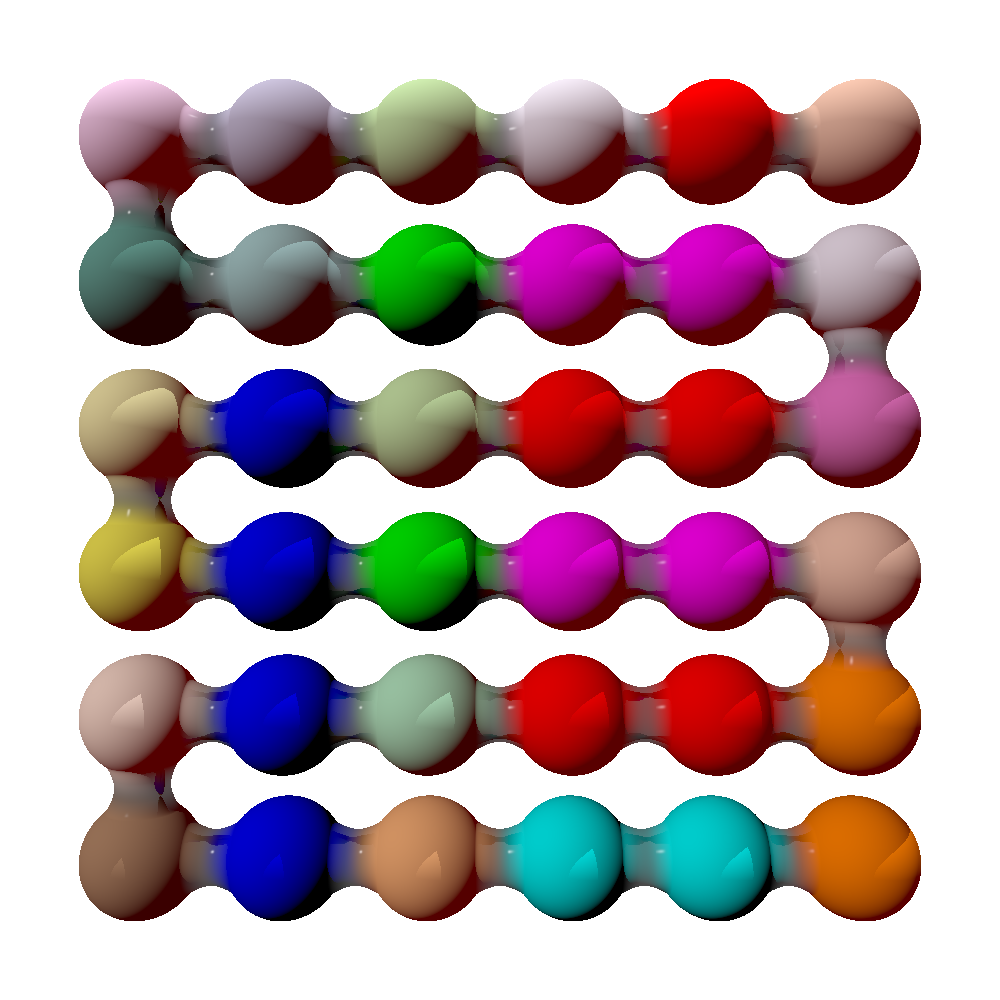}
 \includegraphics[scale=0.07]{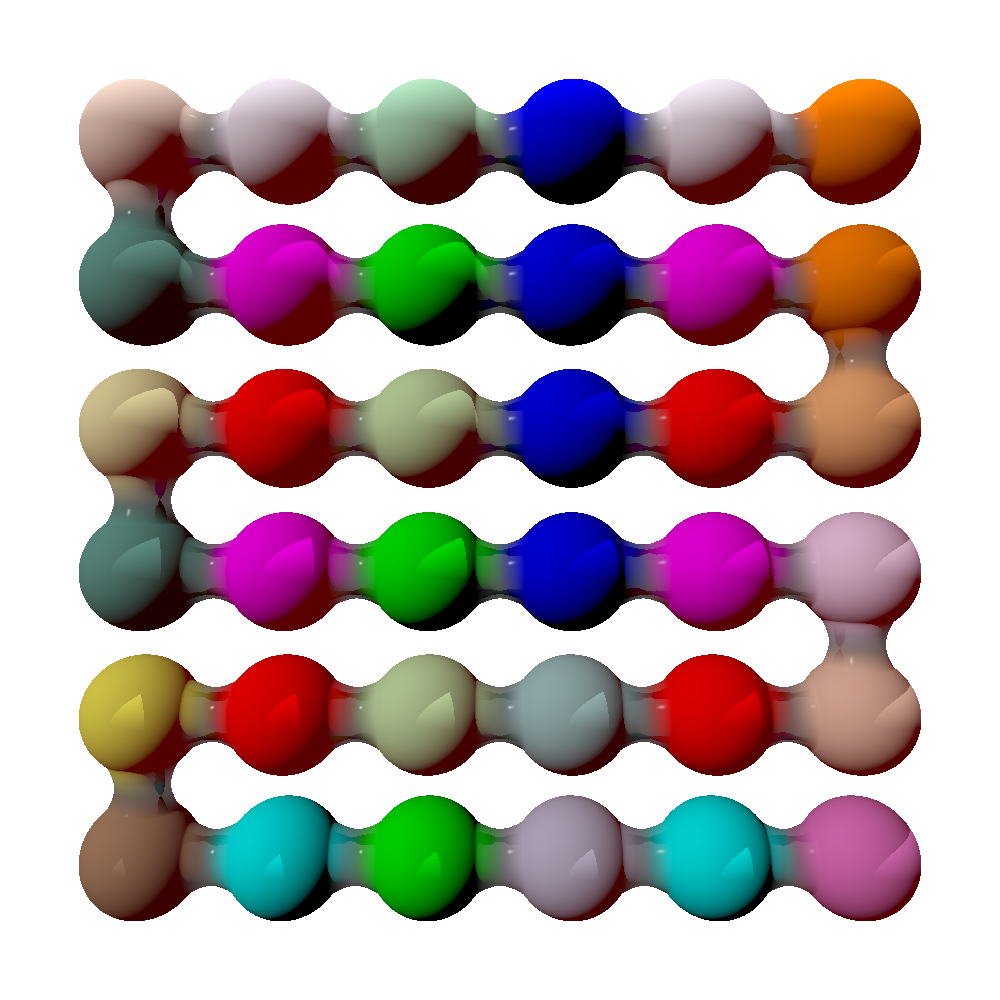}\\
 Protein $B$ \hspace{1 cm} Protein $C$ \hspace{1 cm} Protein $D$\\
 \includegraphics[scale=0.07]{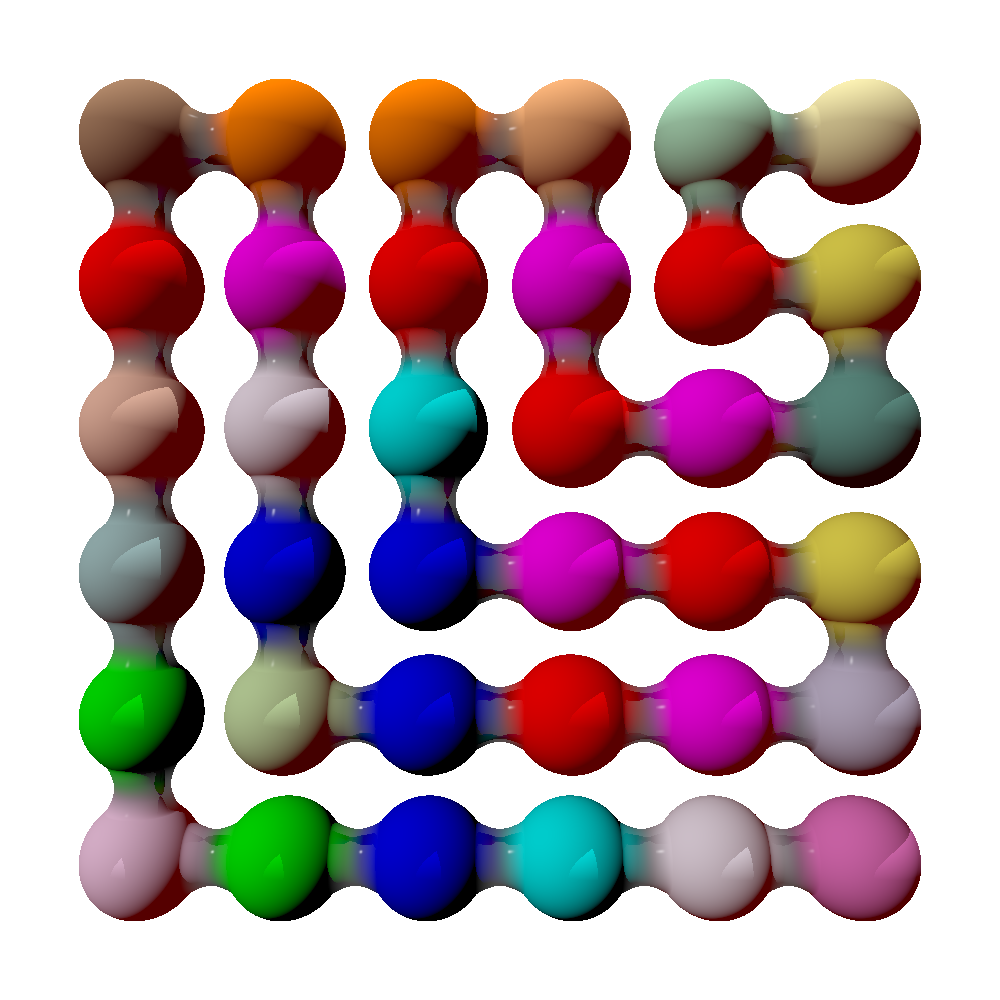}
 \includegraphics[scale=0.07]{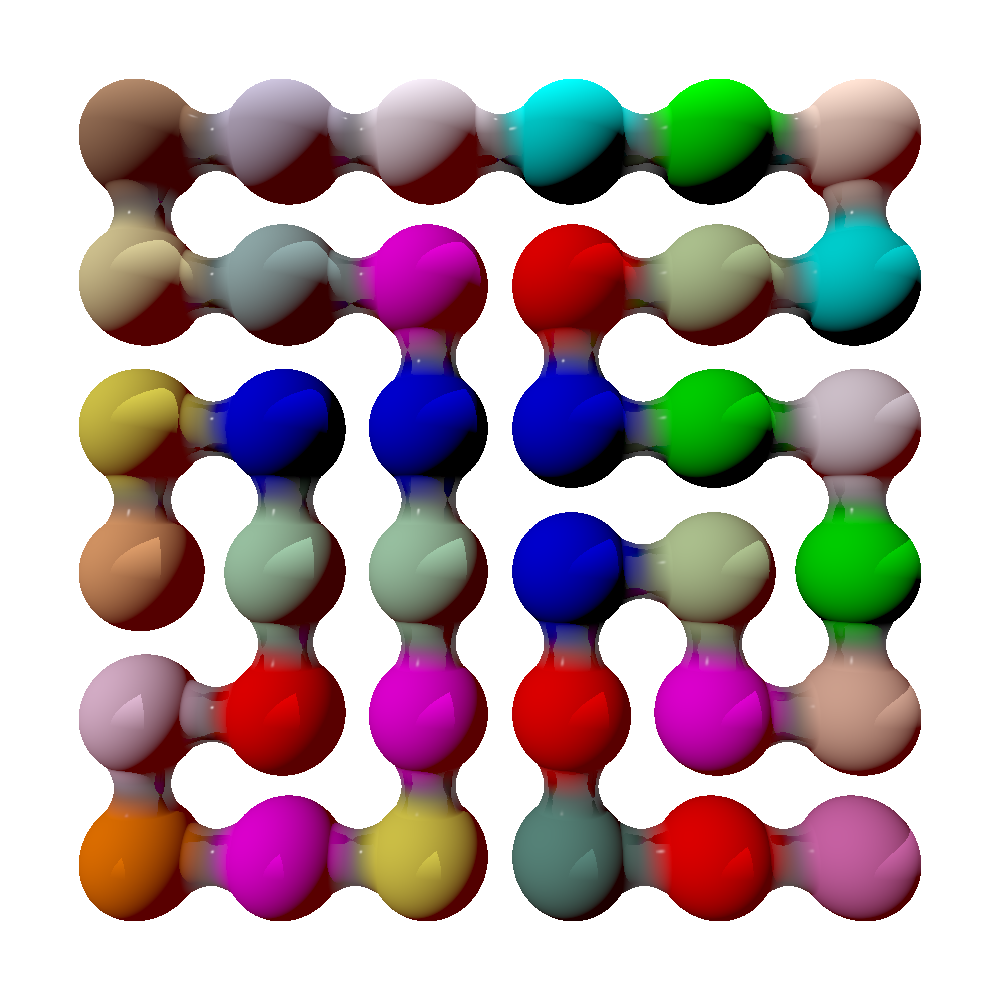}
 \includegraphics[scale=0.07]{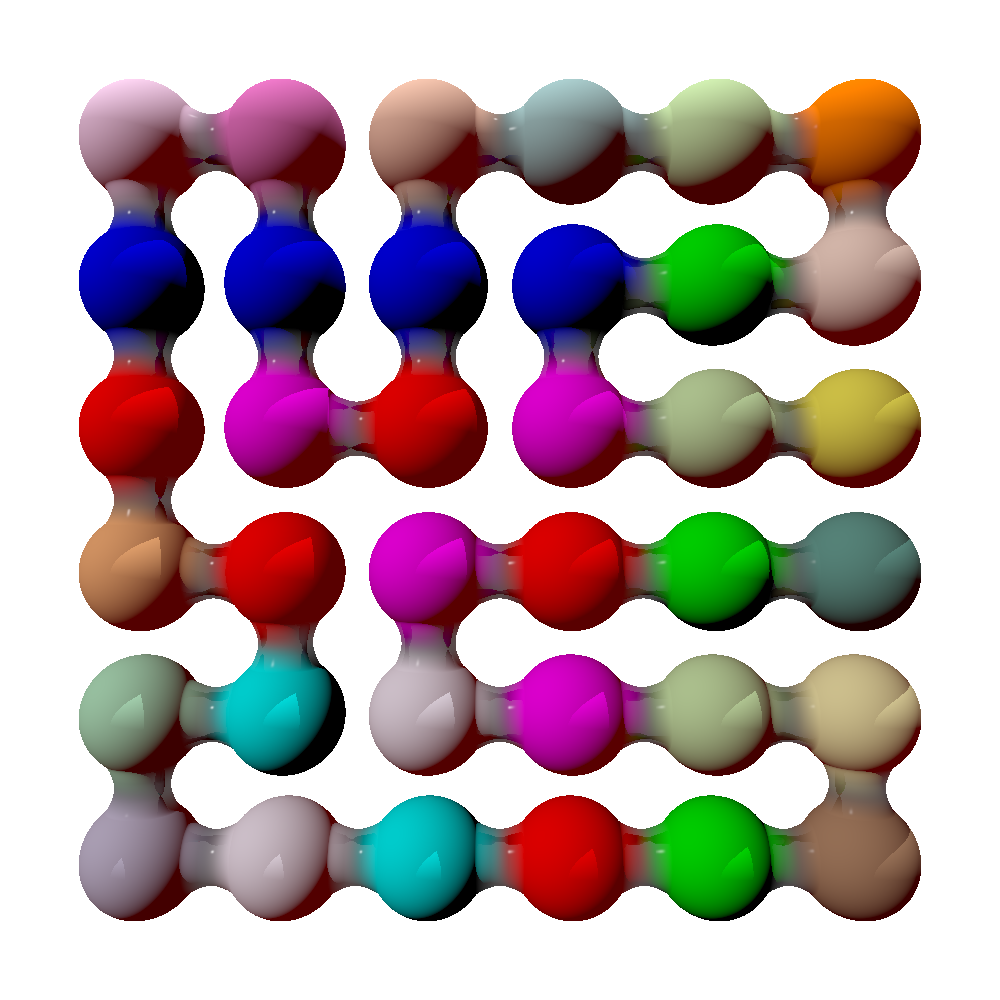}\\
 Protein $E$ \hspace{1 cm} Protein $F$ \\
 \includegraphics[scale=0.07]{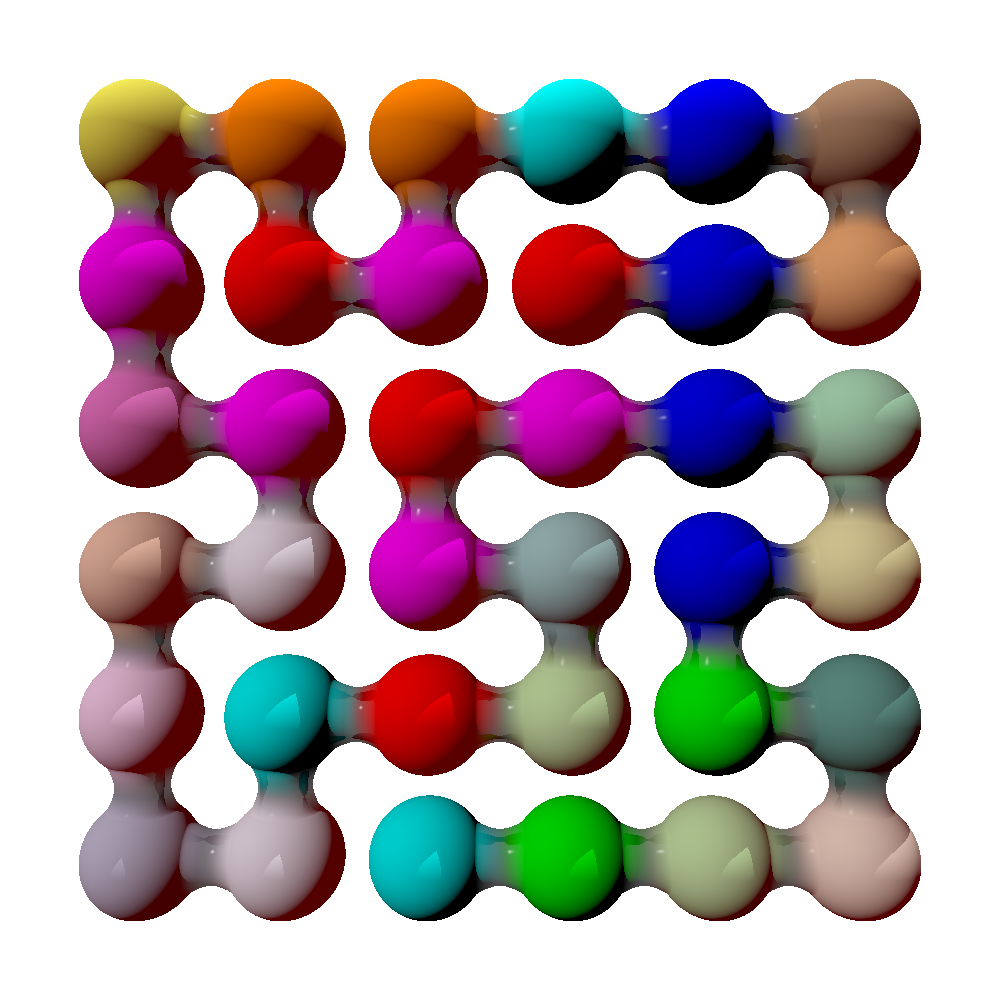}
 \includegraphics[scale=0.07]{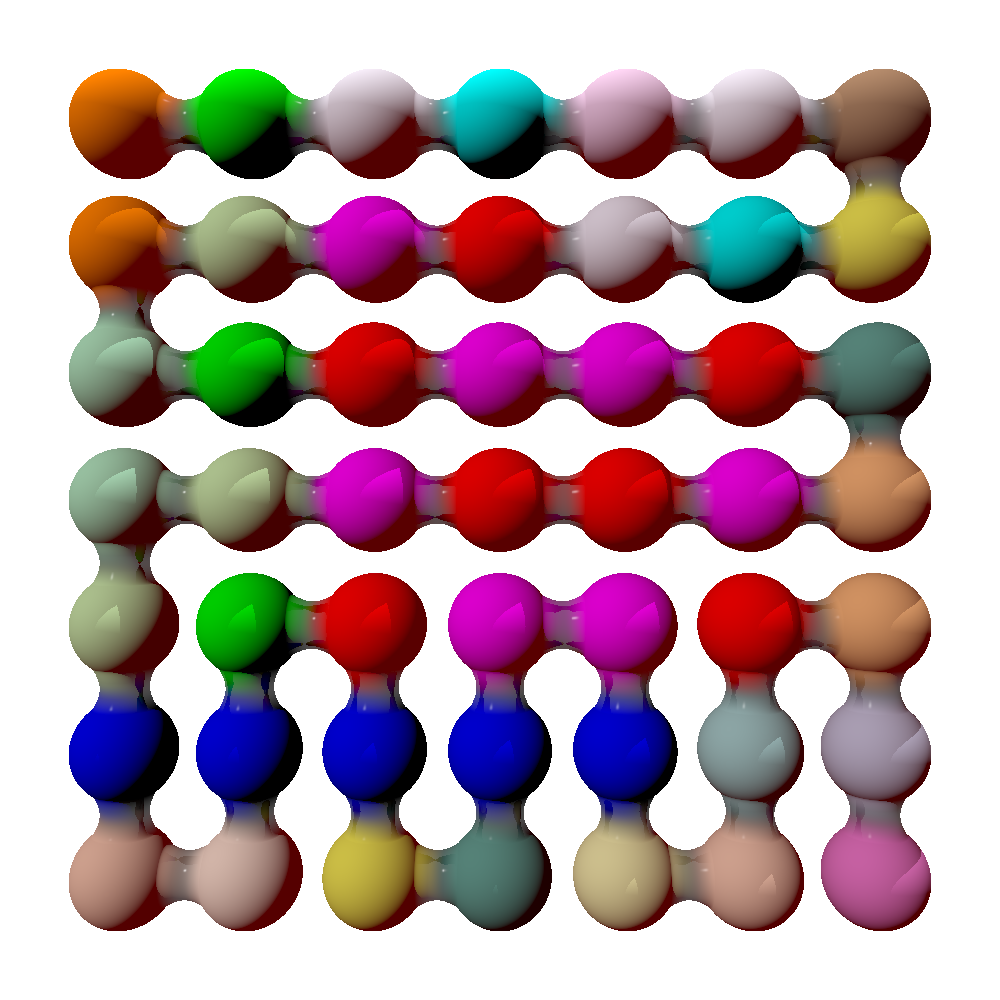}
 \caption{Here we show all the proteins considered in our simulations. Each amino acid is represented with a different color. The sequences of the proteins are the following. Protein 
 $A_0$: $n$ $q$ $m$ $c$ $b$ $t$ $a$ $b$ $i$ $u$ $c$ $n$ $o$ $i$ $m$ $d$ $b$ $s$ $h$ $b$ $i$ $u$ $d$ $p$ $f$ $i$ $m$ $d$ $g$ $l$ $r$ $v$ $i$ $e$ $d$ $e$; 
 Protein 
$A_1$: $n$ $c$ $c$ $f$ $b$ $r$ $a$ $b$ $e$ $i$ $i$ $n$ $p$ $d$ $d$ $m$ $b$ $h$ $t$ $b$ $u$ $i$ $i$ $o$ $q$ $d$ $d$ $m$ $g$ $l$ $s$ $v$ $u$ $q$ $i$ $p$; 
Protein 
$A_2$: $o$ $c$ $v$ $m$ $c$ $r$ $h$ $i$ $u$ $g$ $i$ $p$ $s$ $d$ $b$ $m$ $d$ $l$ $t$ $i$ $u$ $b$ $i$ $f$ $n$ $d$ $b$ $m$ $d$ $l$ $a$ $q$ $e$ $b$ $q$ $n$; 
Protein 
$B$: $o$ $q$ $c$ $b$ $m$ $s$ $m$ $g$ $p$ $i$ $r$ $n$ $d$ $q$ $b$ $u$ $b$ $i$ $d$ $v$ $h$ $i$ $d$ $b$ $c$ $i$ $n$ $f$ $d$ $i$ $d$ $l$ $h$ $i$ $e$ $t$; 
Protein 
$C$: $o$ $i$ $l$ $i$ $b$ $u$ $d$ $p$ $m$ $q$ $m$ $b$ $i$ $u$ $c$ $a$ $m$ $c$ $q$ $v$ $r$ $t$ $g$ $d$ $b$ $e$ $d$ $h$ $d$ $n$ $s$ $i$ $e$ $b$ $h$ $f$;
Protein 
$D$: $l$ $m$ $i$ $d$ $q$ $d$ $u$ $t$ $r$ $m$ $i$ $c$ $q$ $v$ $e$ $c$ $i$ $f$ $i$ $b$ $s$ $o$ $b$ $d$ $i$ $b$ $p$ $g$ $u$ $n$ $a$ $m$ $b$ $d$ $u$ $h$; 
Protein 
$E$: $c$ $m$ $u$ $a$ $l$ $m$ $b$ $t$ $e$ $b$ $d$ $i$ $d$ $g$ $u$ $i$ $c$ $q$ $v$ $s$ $p$ $q$ $d$ $o$ $d$ $h$ $n$ $i$ $d$ $n$ $c$ $b$ $r$ $f$ $b$ $i$; 
Protein 
$F$: $o$ $v$ $f$ $i$ $g$ $p$ $t$ $b$ $d$ $d$ $b$ $l$ $h$ $b$ $i$ $m$ $b$ $a$ $p$ $b$ $u$ $e$ $u$ $d$ $i$ $i$ $d$ $f$ $l$ $i$ $d$ $d$ $i$ $m$ $e$ $n$ $u$ $d$ $i$ $q$ $c$ $h$ $r$ $q$ $s$ $c$ $q$ $m$ $n$.
By shifting one sequence with respect to the other we establish the maximum overlap between them. We find that $A_0$ and $A_1$ have 10 amino acids in the same position 
; $A_0$ and $A_2$ have 6 corresponding amino acids
; $A_0$ and $B$ have 5 overlapping amino acids
; $A_0$ and $C$ share 5 amino acids
; $A_0$ and $D$ have 6 amino acids in common
;  $A_0$ and $E$ have 8 amino acids in common
; $A_0$ and $F$ share 6 amino acids
; $B$ and $C$ share 6 amino acids
.
 }
 \label{protein}
\end{figure}

\begin{figure}
\centering
 \includegraphics[scale=0.4]{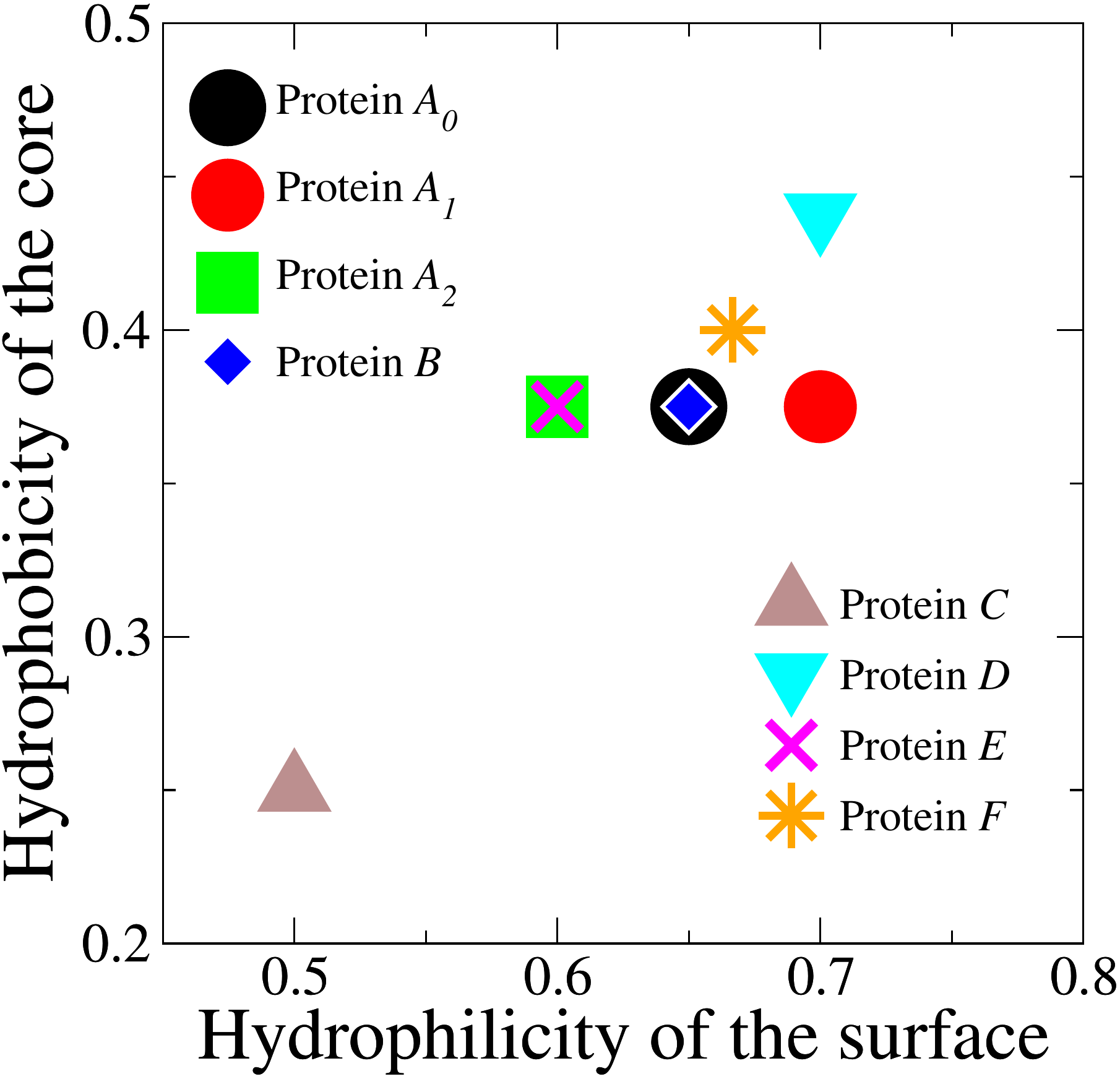}
 \caption{Composition of the designed proteins. The hydrophilicity (hydrophobicity) of the protein surface (core) is given by the ratio between the number of hydrophilic (hydrophobic) amino acids on the surface (core) and the total number of amino acids exposed to the solvent (buried into the core) when the protein attains its native conformation. }
 \label{composition}
\end{figure}

\begin{figure}
 \includegraphics[width=\textwidth]{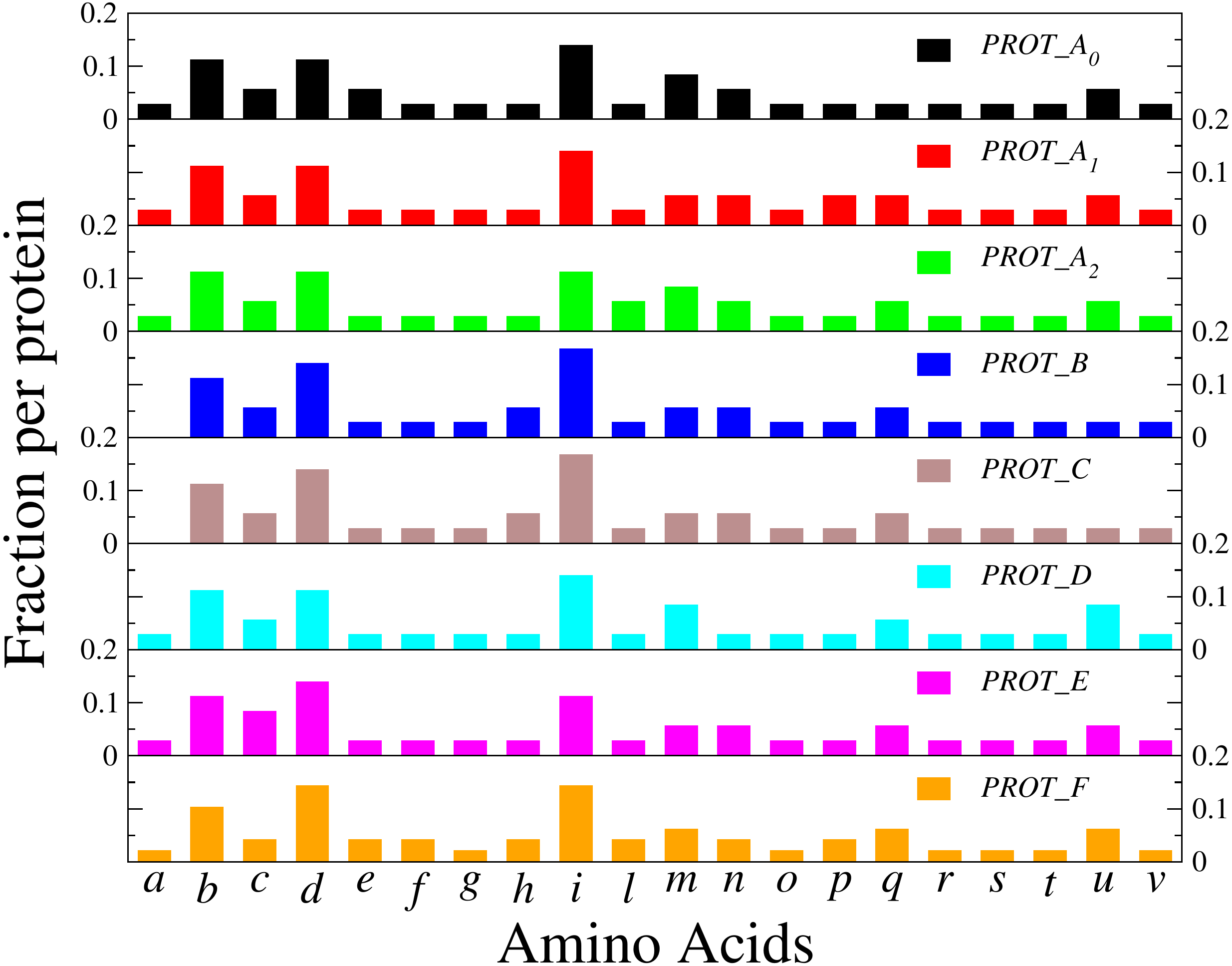}
 \caption{Amino acid composition of the designed proteins. }
 \label{sequences}
\end{figure}

We consider 8 different proteins, which we label as $A_0$, $A_1$, $A_2$, $B$, $C$, $D$, $E$ and $F$, which native states are shown in Fig. \ref{protein}. Each capital letter in the protein label identifies a different native structure, while different subscript numbers refer to different sequences associated to the structure. Therefore, proteins $A_0$, $A_1$ and $A_2$ share the same native structure, but have a different sequences.
All the native structures have been selected considering maximally compact conformations, composed by 36 or 49 amino acids. Then, for each native structure, the protein sequence has been established through a design scheme, based on the standard approach introduced by Shakhnovich and Gutin \cite{Shakhnovich1993c, Shakhnovich1993a} and successfully adopted to design realistic off-lattice proteins \cite{Cardelli2017, Cardelli2018, Cardelli2019, Nerattini2019},  but
accounting explicitly for the water properties in the protein hydration shell \cite{Bianco2017}. 
We perform a Monte Carlo simulations in the isobaric--isothermal ensemble at ambient conditions, keeping fixed the protein conformation in its native state and mutating the amino acids, to explore the phase space of sequences. For each sequence the surrounding water is equilibrated and the average enthalpy $H$ of the hydrated protein (residue--residue energy plus the average enthalpy of the water molecules in the hydration shell) is computed. The sequence to whom correspond the minimum value of $H$ is selected as best folder.
The design scheme leads to sequences which are not perfectly hydrophilic on the surface and hydrophobic into the core, consistent with what is observed in real proteins \cite{Lins2003,Moelbert2004}. The hydropathy of the designed protein surface and core is shown in Fig. \ref{composition}, while the full amino acid composition composition of each sequence is shown in Fig. \ref{sequences}. It is worth to be noted that all the sequences generated differ each other, exhibiting different values for the hydrophilicity (hydrophobicity) of the protein surface (core), irrespective of the native structure, and the maximum overlap between the sequences is of 10 amino acids\footnote{The maximum overlap between two sequences is computed shifting and overlapping one sequence with respect to the other, and counting the number of amino acids on both sequences which coincides along the overlapped region.}. 
Each designed sequence is folded alone at ambient conditions to prove its capability to reach the native state.  
Once the proteins have been designed, we simulate the folding of multi-protein systems in a range of concentrations $c\in[1\%,55\%]$, considering homogeneous solutions, i.e. when all the sequences are equal. Along the simulations we compute the free energy landscape as function of the total number of native contacts $N_c$ and inter-protein contacts $I_c$ to study the folding--aggregation competition.

\section{Water Model}

The coarse-grain representation of the water molecules replaces the  coordinates and orientations of the water molecules by a continuous density field and discrete bonding variables, respectively. The density field is defined on top of a partition of the volume  $V$ into a fixed number $N$ of cells, each with volume $v\equiv V/N\geq v_0$, being $v_0\equiv r_0^3$ the water excluded volume and $r_0\equiv 2.9 $ \AA{}  the water van der Waals diameter. 
The size of a cell $r\ge r_0$ is a stochastic variable and coincides, by construction, with the average distance between first-neighbour water molecules. The general formulation of the model envisages to each cell $i$ an index $n_i = 1$ or $n_i = 0$ according to the size $r$ (which varies a lot from the gas phase to the super-cooled one), to distinguish when the molecule can form hydrogen bonds (HBs) or not, respectively. Here, since we perform the study at ambient conditions, we assume that all the molecules can form HB, placing $n_i=1$ to all cells, therefore such an index is removed from the following expression for sake of simplicity (for general formulation see for example Ref. \cite{Stokely2010, Mazza2011, delosSantos2011, PhysRevLett.109.105701, Franzese2013, Bianco2014,Franzese2011, Bianco2012a, Mazza19873, Bianco2015, Bianco2017a, Bianco2017, Bianco2019}).

The Hamiltonian of the bulk water is
\begin{equation}
 \mathscr{H}_{\rm w,w}\equiv \sum_{ij} U(r_{ij}) -J N_{\rm HB}^{\rm (b)}-J_\sigma N_{\rm  coop}^{\rm (b)}.
\label{bulk}
\end{equation}
The first term, summed over all the water molecules $i$ and $j$ at oxygen-oxygen distance $r_{ij}$, is given by $4\epsilon[(r_0/r)^{12} - (r_0/r)^6]$ for $r_0<r<6r_0$, $U=\infty$ for $r\le r_0$, and $U=0$ for $r\ge 6r_0$ (cutoff distance). We fix $\epsilon$ = 2.9 kJ/mol. 

The second term of the Hamiltonian represents the directional component of the water-water hydrogen bonds (HB). By assuming that a molecule can form up to four HBs, we discretize the number of possible molecular conformations introducing four bonding indices $\sigma_{ij}$ for each water molecule $i$. the variable $\sigma_{ij}$ describes the bonding conformation of the molecule $i$ with respect to the neighbour molecule $j$. Each variable $\sigma_{ij}$ has $q$ possible states, and if $\sigma_{ij}=\sigma_{ji}$ an HB between the molecules $i$ and $j$ is formed, with the characteristic energy $J/4\epsilon=0.3$. The number of HB is then defined as $N^{\rm(b)}_{\rm HB}\equiv \sum_{\langle ij \rangle} \delta_{\sigma_{ij},\sigma_{ji}}$, with $\delta_{ab}=1$ if $a=b$, 0 otherwise.
A HB is broken if the oxygen-oxygen-hydrogen angle exceeds the 30$^\circ$, therefore only 1/6 of the entire range of values [0,360$^\circ$] of this angle is associated to a bonded state. Fixing $q=6$ we correctly account for the entropic loss due to the HB formation. 

The third interaction term in Eq. (\ref{bulk}) corresponds to the cooperative interaction of the HBs due to the oxygen-oxygen-oxygen correlation. This effect originates from quantum many-body interactions of the HB \cite{doi:10.1021/ja0424676} and in the bulk leads the molecules toward an ordered tetrahedral configuration \cite{PhysRevLett.84.2881}. This term is modelled as an effective interaction--with coupling constant $J_\sigma$--between each of the six different pairs of the four indexes $\sigma_{ij}$ of a molecule $i$. Hence, we have $N_{\rm  coop}^{\rm (b)}\equiv  \sum_{ikl}\delta_{\sigma_{ik},\sigma_{il}}$ which defines the cooperativity of the water molecules. By assuming $J_\sigma \ll J$  we guarantee the asymmetry between the two terms \cite{Stokely2010}.

For any HB formed in the bulk the local volume increases of the quantity $v^{(b)}_{HB}/v_0$. The associated enthalpic variation is $-J+Pv^{(b)}_{HB}$, being $P$ the pressure. It accounts for the $P$ disrupting effect on the HB network. Here $v^{(b)}_{HB}/v_0$ represents the average volume increase between
high-density ices VI and VIII and low-density ice Ih \cite{Stokely2010}. Hence, the volume of bulk molecules is given by $V^{(b)}=Nv + N^{(b)}_{\rm HB} v^{(b)}_{\rm HB}.$

The water-water hydrogen bonding in the protein hydration shell depends on the hydrophobic (\texttt{PHO}) or hydrophilic (\texttt{PHI})  nature of the hydrated amino acids, and is described by the Hamiltonian  
\begin{eqnarray}
 \mathscr{H}_{\rm w,w}^{(h)}  \equiv  -  \left[  J^{\texttt{PHO}} N_{\rm HB}^{\rm \texttt{PHO}}+J^{\texttt{PHI}} N_{\rm HB}^{\rm \texttt{PHI}} + J^{\texttt{MIX}} N_{\rm HB}^{\rm \texttt{MIX}}\right] + \\\nonumber -  \left[
 J^{\texttt{PHO}}_\sigma N_{\rm coop}^{\rm \texttt{PHO} }+J^{\texttt{PHI}}_\sigma N_{\rm coop}^{\rm \texttt{PHI} } + J^{\texttt{MIX}}_\sigma N_{\rm coop}^{\rm \texttt{MIX} }\right] , \label{water}
\end{eqnarray}
where $N_{\rm HB}^{\rm \texttt{PHO}}$, $N_{\rm HB}^{\rm \texttt{PHI}}$ and $N_{\rm HB}^{\rm \texttt{MIX}}$ indicate respectively the number of HB formed between two molecules hydrating two hydrophobic amino acids, two hydrophilic amino acids, one hydrophobic amino acid and one hydrophilic amino acid. Analogously $N_{\rm coop}^{\rm \texttt{PHO}}$, $N_{\rm coop}^{\rm \texttt{PHI}}$ and $N_{\rm coop}^{\rm \texttt{MIX}}$ represent the cooperative bonds at the hydrophobic, hydrophilic and mixed interface. 

The hydrophobic interface strengthens the water-water hydrogen bonding in the first hydration shell \cite{GiovanbattistaPRE2006, DiasPRL2008, DavisNat2012, SarupriaPRL2009} and increases the local water density upon pressurization \cite{ SarupriaPRL2009, DasJPCB2012, GhoshJACS2001,  DiasJPCB2014}. The first effect is included by assuming $J^{\texttt{PHO}}>J$ and $J_\sigma^{\texttt{PHO}} > J_\sigma$. This condition guarantees that the solvation free energy  of a hydrophobic amino acid decreases at low temperature $T$ \cite{moghaddam2007}. The second one is accounted assuming that the volume associate to the HB at the \texttt{PHO} interface decreases upon increasing $P$, $v_{\rm  HB}^{\texttt{PHO}}/v_{\rm HB,0}^{\texttt{PHO}}\equiv 1- k_1P $ \cite{Bianco2015}. In this way, the density fluctuations at the \texttt{PHO} interface are reduced at high $P$. The volume contribution $V^{\texttt{PHO}}$ to total volume $V$ due the HBs in the hydrophobic shell is $ V^{\texttt{PHO}} \equiv N_{\rm HB}^{\texttt{PHO}} v_{\rm  HB}^{\texttt{PHO}}$. We assume that the water-water hydrogen bonding and the water density at the hydrophilic interface are not affected by the protein. Therefore, $J^{\texttt{PHI}}=J$, $J_\sigma^{\texttt{PHI}}=J_\sigma$ and $v_{\rm HB}^{\texttt{PHI}} = v_{\rm HB}^{(b)}$. Finally, we fix $J^{\texttt{MIX}} \equiv (J^{\texttt{PHO}} +   J^{\texttt{PHI}})/2 $ and $J^{\texttt{MIX}}_\sigma \equiv (J^{\texttt{PHO}}_\sigma +   J^{\texttt{PHI}}_\sigma)/2$.

Lastly, we assume that the protein-water interaction energy is $-\varepsilon^{\texttt{PHO}}$ or $- \varepsilon^{\texttt{PHI}}$, depending if the residue is hydrophobic or hydrophilic, respectively. As reported in Ref. \cite{Bianco2017}, we express the model parameters in units of $8\epsilon$, and fix the value to 
$J = 0.3$ and $J_\sigma =0.05$ (bulk water), $J^{\texttt{PHI}} = J$ and $J_\sigma^{\texttt{PHI}}=J_\sigma $ (water at hydrophilic interfaces), $J^{\texttt{PHO}}=1.2$ and $J^{\texttt{PHO}}_\sigma=0.2$ (water at hydrophobic interfaces), $\varepsilon^{\texttt{PHO}} = 0$ or $ \varepsilon^{\texttt{PHI}} = 0.48$. 
Finally, we fix $k_1=4$, $v_{\rm HB}^{(b)}/v_0=0.5$ and   $v_{\rm HB, 0}^{\texttt{PHO}}/v_0=2$. 
These choices balance the water-water, the water-residue and the residue-residue interactions, making the proteins stable for thermodynamic conditions comprised in the (stable and metastable) liquid phase, including ambient conditions. Moreover, by enhancing the interface interactions we account for the lower surface volume ratio of the model (formulated in two dimensions) with respect to a three-dimensional system. 

All the results presented in this work have been tested under the change of parameters. In particular, we have decreased the effect of the protein interface on the water-water interaction observing a decrease in the concentration thresholds at which the proteins unfold and aggregate, but the phenomenology observed is substantially the same.

\section{Folding vs Aggregation in homogeneous protein solutions}

\begin{figure}
\centering
{\bf Protein $A_0$}\\
\includegraphics[scale=0.53]{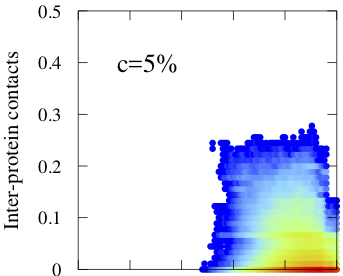}
\includegraphics[scale=0.53]{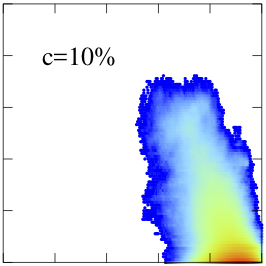}
\includegraphics[scale=0.53]{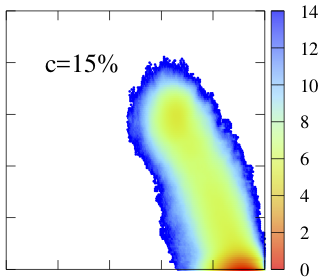}\\
\includegraphics[scale=0.53]{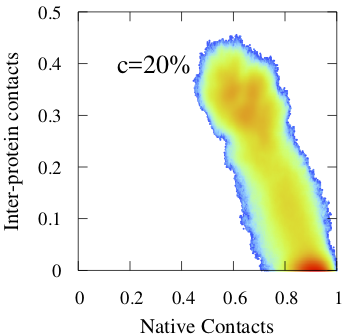}
\includegraphics[scale=0.53]{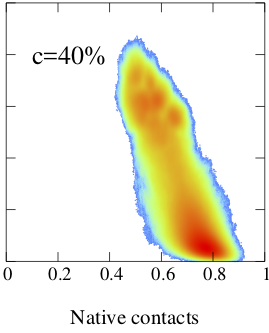}
\includegraphics[scale=0.53]{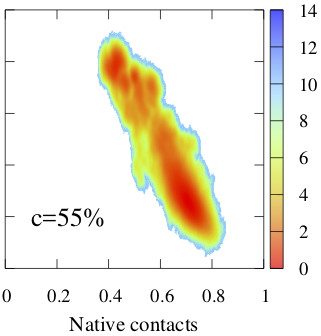}\\
 {\bf Protein $B$}\\
\includegraphics[scale=0.53]{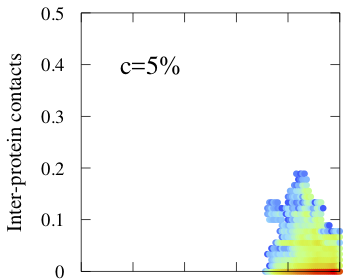}
\includegraphics[scale=0.53]{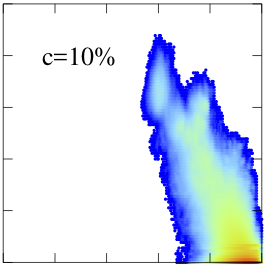}
\includegraphics[scale=0.53]{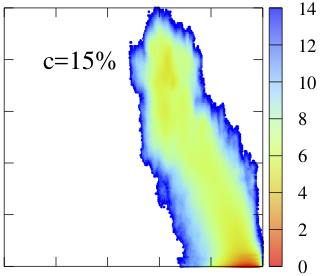}\\
\includegraphics[scale=0.53]{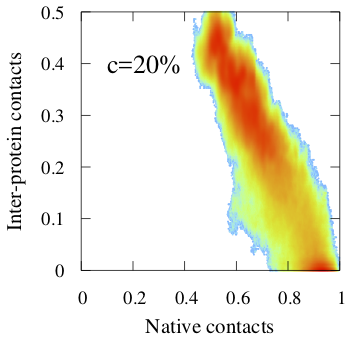}
\includegraphics[scale=0.53]{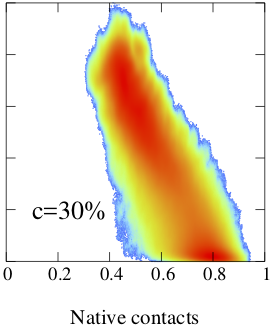}
\includegraphics[scale=0.53]{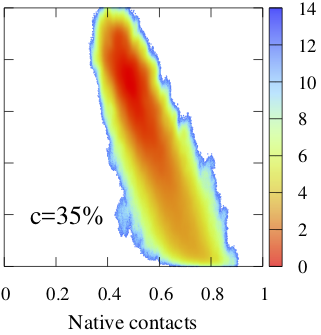}\\
 {\bf Protein $C$}\\
\includegraphics[scale=0.53]{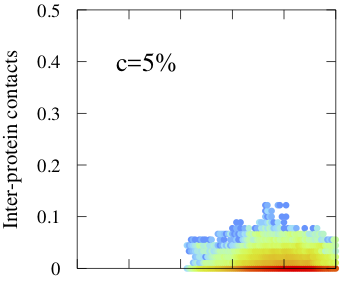}
\includegraphics[scale=0.53]{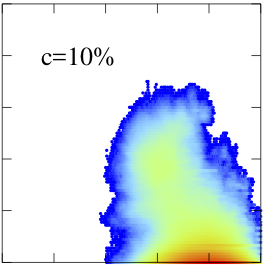}
\includegraphics[scale=0.53]{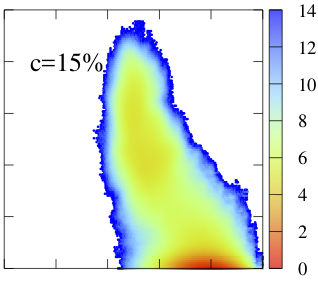}\\
\includegraphics[scale=0.53]{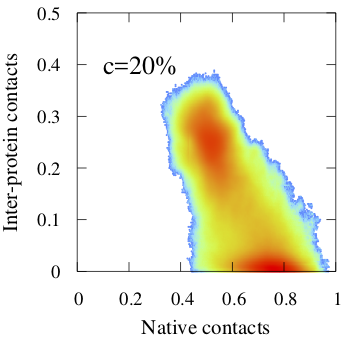}
\includegraphics[scale=0.53]{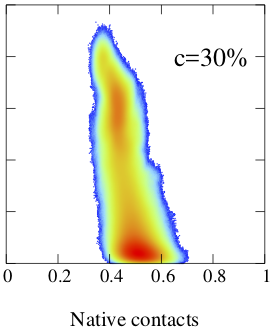}
\includegraphics[scale=0.53]{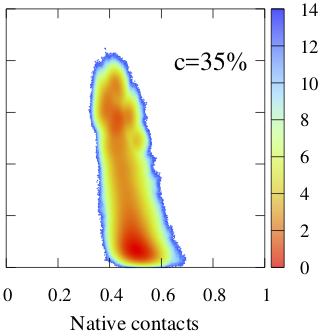}\\
\caption{Color map of the free energy profile of the protein $A_0$, $B$ and $C$, as function of the native contacts and inter-protein contacts, for different protein concentration $c$. Native contacts have been normalized to 1 and inter-protein contacts have been normalized to $ln$, where $n$ is the number of proteins simulated and $l$ is the length (number of amino acids) of a single protein. In the shown cases, the size of the simulation box have been chosen such that $c=n$, i.e. a single protein occupies a volume corresponding to the 1\% of the available volume.}
 \label{F_Nc_Ic}
\end{figure}

\begin{figure}
\centering
{\bf (a)\hspace{5.2 cm} (b)} \\ 
 \begin{overpic}[scale=0.21]{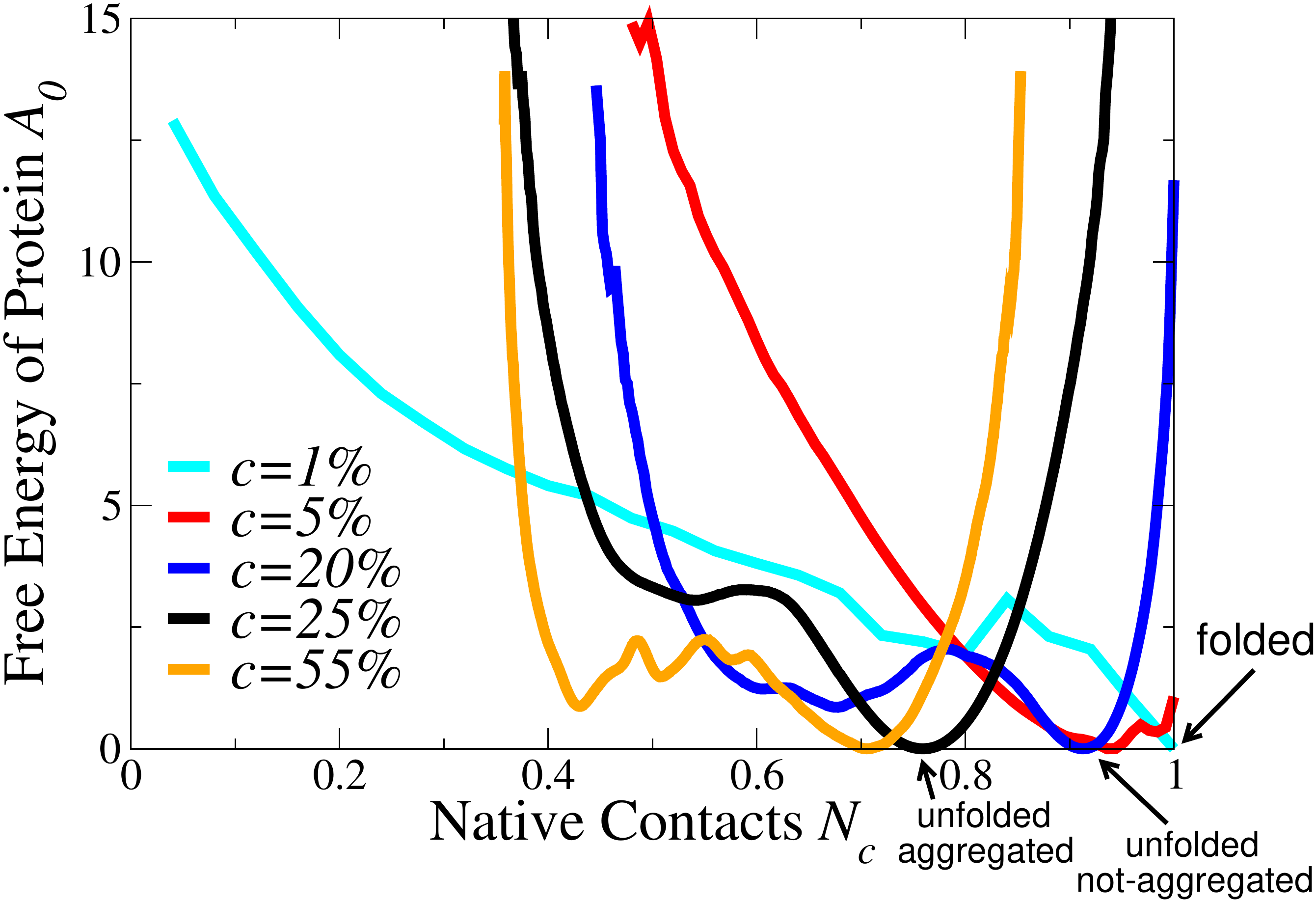}
 \put(11,41){\includegraphics[scale=0.04]{prot_A0.png}}
\end{overpic}
\begin{overpic}[scale=0.21]{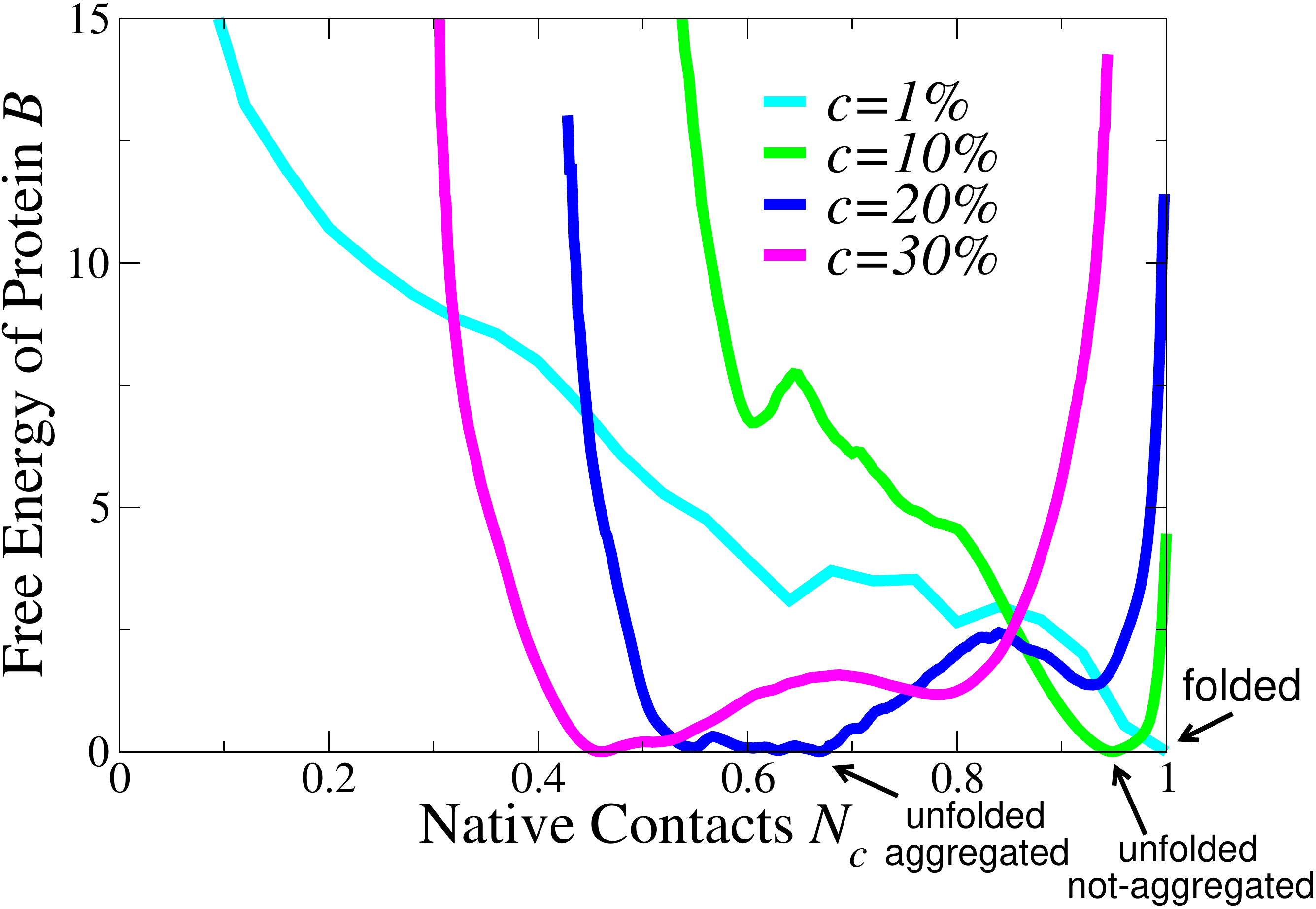}
 \put(10,13){\includegraphics[scale=0.04]{prot_B.png}}
\end{overpic} \\
{\bf (c)\hspace{5.2 cm} (d)}
\begin{overpic}[scale=0.21]{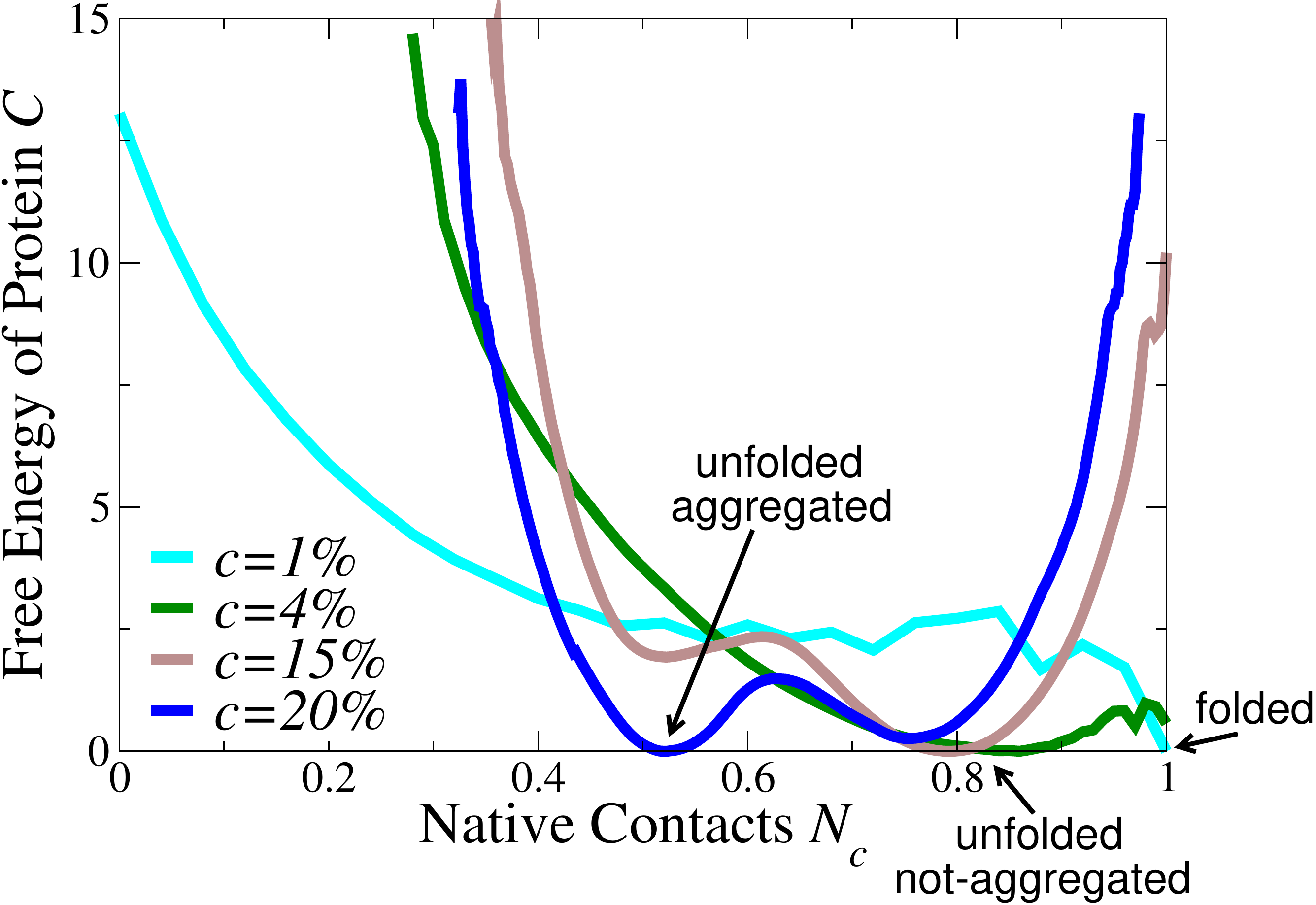}
 \put(54,42){\includegraphics[scale=0.04]{prot_C.png}}
\end{overpic}
\includegraphics[scale=0.21]{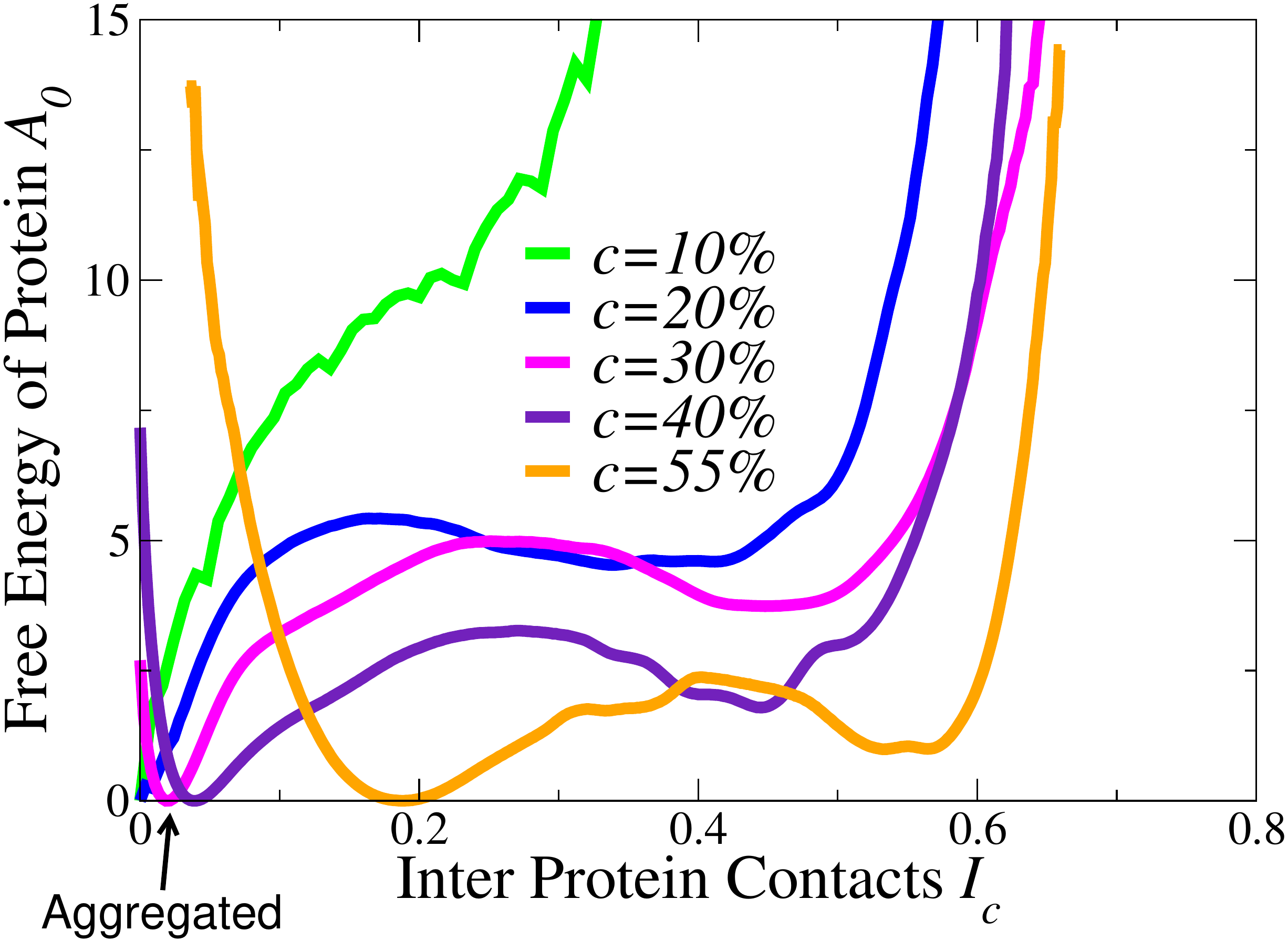}\hspace{0.45 cm}\\
{\bf (e) \hspace{5.1 cm} (f) }\\
\includegraphics[scale=0.21]{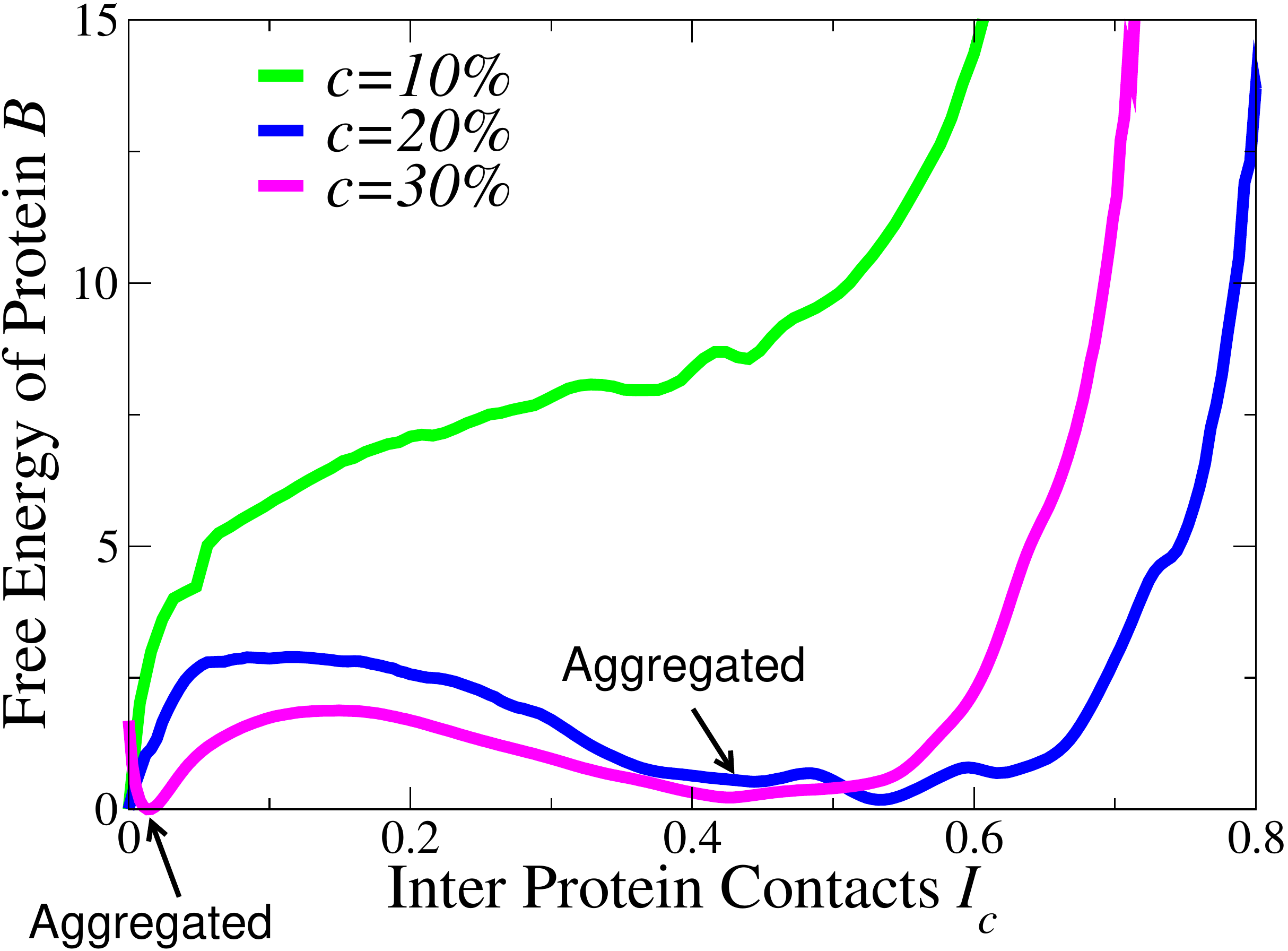}\hspace{0.45 cm}
\includegraphics[scale=0.21]{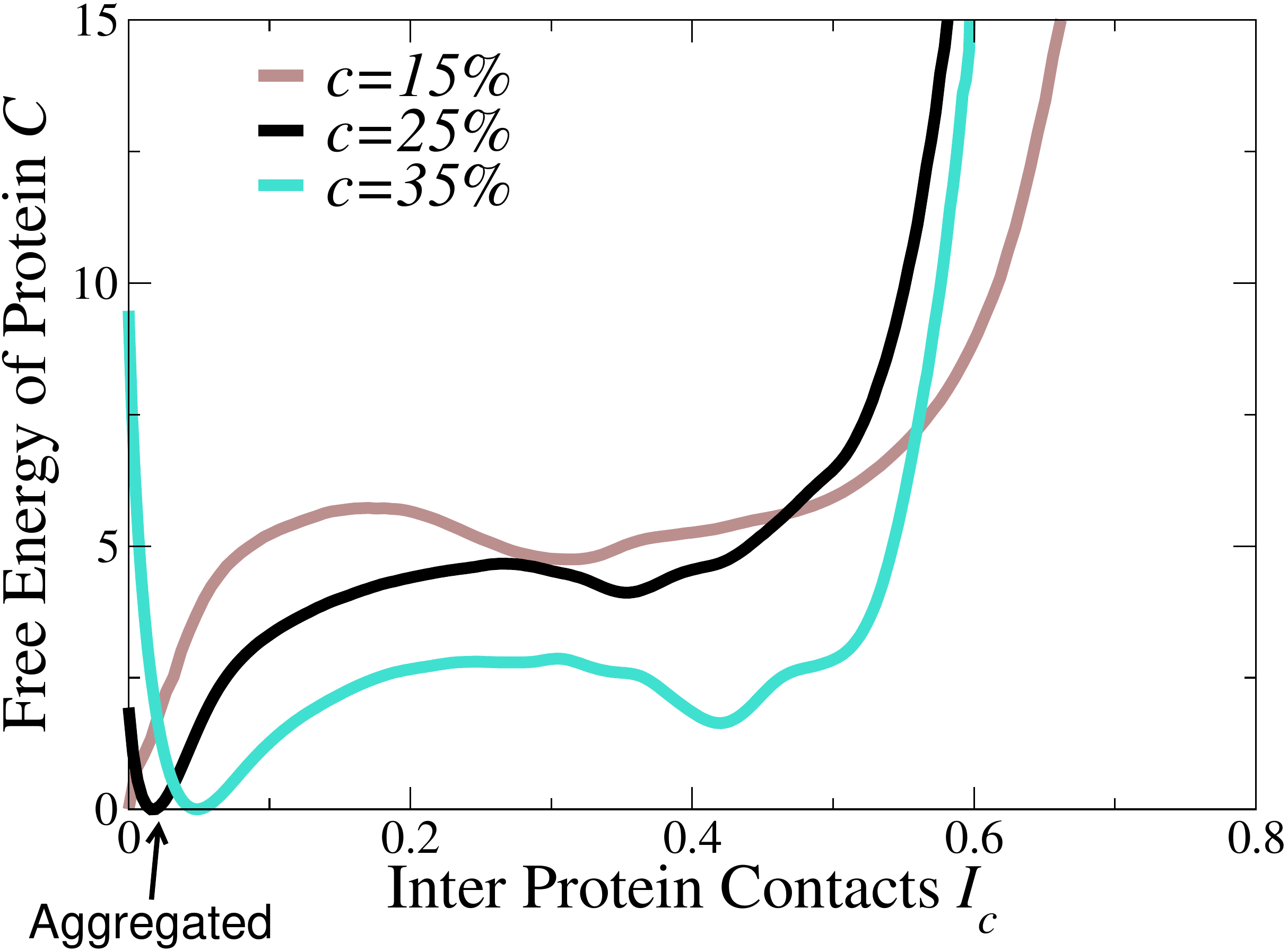}
\caption{Free energy profile of the protein $A_0$, $B$ and $C$ as function of $N_c$ (upper panels) and $I_c$ (lower panels) for different concentrations. All the free energy curves are in $k_BT$ units and have been shifted such that the minimum coincides with 0. The $N_c$ axes has been normalized dividing the  number of native contacts for its maximum possible value (corresponding to all the proteins in their native conformation). The axes $Ic$ has been normalized dividing the number of inter-protein contacts for the total number of amino acids. 
We find that, for the protein $B$ ($C$), the $FOL\rightarrow UNF$ transition  occurs at $c_{FOL\rightarrow UNF}^{(B)} \sim 8\pm 1 \%$ ($c_{FOL\rightarrow UNF}^{(C)} \sim 4\pm 1 \%$) and the $UNF\rightarrow AGG$ transition  occurs at $c_{UNF\rightarrow AGG}^{(B)} \sim 16.5\pm 1.5 \%$ ($c_{UNF\rightarrow AGG}^{(C)} \sim 18\pm 2 \%$). }
\label{F_Nc}
\end{figure}

In Fig. \ref{F_Nc_Ic} we show the free energy landscape of proteins $A_0$, $B$ and $C$ as function of $N_c$ and $I_c$ simulated in a concentration range $c\in[1\%,55\%]$. 
In all the cases we observe that for low concentrations, $c\lesssim 5\%$, the minimum of the free energy correspond to $N_c = 1$ and $I_c = 0$, i.e. all the proteins reach their native folded state and, on average, are not in contact to each other. 

By looking at the separate free energy profile as function of $N_c$ (Fig. \ref{F_Nc}a,b,c) and $I_c$ (Fig. \ref{F_Nc}d,e,f) (obteined integrating the free energy profiles shown if Fig. \ref{F_Nc_Ic} along the axes $I_c$ and $N_c$ respectively), respectively indicated with $F(N_c)$ and $F(I_c)$, we can identify three different states for each protein: i) the native state $FOL$; ii) the unfolded and not aggregated state $UNF$; iii) the unfolded and aggregated state $AGG$. The $FOL$ state occurs when  all the proteins recover their native conformation and the minima $F_{\rm min}(N_c)$ and $F_{\rm min}(I_c)$, respectively of the free energy profiles $F(N_c)$ and $F(I_c)$, occur at $N_c=1$ and $I_c=0$. The unfolded and not-aggregated  state $UNF$ takes place when the protein looses part of its native contacts leading to $F_{\rm min}(N_c)$ for  $0.8 \lesssim N_c < 1$ while the aggregated state is still less favourable being $F_{\rm min}(I_c)$ for $I_c=0$. 
The peculiar characteristic of the $UNF$ state is that there are no inter-protein contacts ($I_c=0$ remains by far the lowest free energy minima Fig.\ref{F_Nc}b).

In Fig. \ref{fig:folding_vs_distance} we prove that, for protein isolated pairs, the unfolding starts before the residues can interact directly.
Since the proteins are not interacting directly, and there are no long-range interactions in the model the logical conclusion is that the water is mediating the interaction that stabilises the misfolded states compared to the folded one. When we switched off the water terms in the model the $UNF$ state disappears, and the systems go directly into the $AGG$ state at even lower concentrations $c$ (see Fig. \ref{fig:implicit_solvent} in the Supplementary Information). Hence, it is clear that the water is creating a barrier against aggregation.

Such an unexpected role of the water has to the best of our knowledge never been observed before. 

The $UNF$ state holds for quite large values of $c$, where protein gradually unfold by increasing $c$. Eventually, at very high concentrations ($c\ge 20\%$ for protein $A_0$), we observe the appearance of a clear minimum in the free energy ($I_c>0$ in Fig.\ref{F_Nc_Ic}.d-f) signifying that we reached an aggregated state $AGG$.
The occurrence of aggregates $AGG$ comes with a loss of the native conformations ($F_{\rm min}(N_c)|_{N_c<0.8}$) consistent with previous observations \cite{Bratko2001}.

It is important to stress that the concentration thresholds of the $FOL\rightarrow UNF$ and $UNF\rightarrow AGG$ transitions, which we indicate with symbols $c_{FOL\rightarrow UNF}^{(i)}$ and $c_{UNF\rightarrow AGG}^{(i)}$ with $i=A_0, B, C$, depend on the specific sequence (Fig. \ref{F_Nc}). 

By comparing the $UNF\rightarrow AGG$ transition points for proteins $A_0$ and $B$ (Fig. \ref{F_Nc}d,e), which have the same fraction of hydrophilic amino acids on the surface and hydrophobic amino acids into the core (Fig. \ref{composition}), we observe that the protein $A_0$ is less prone to aggregate with respect to $B$, since $c_{UNF\rightarrow AGG}^{(A_0)} > c_{UNF\rightarrow AGG}^{(B)}$. On the other hand, by comparing the same transition points between proteins $B$ and $C$ (Fig. \ref{F_Nc}b,c), we find that both transitions occur at similar values of $c$ within the numerical error, although their surface and core composition are quite different, being the protein $C$ more hydrophobic on the surface and less hydrophobic into the core with respect to the protein $B$. This interesting result points out that the propensity to aggregate of proteins is not strictly related to the hydrophobic content of its surface, as long as this amount has been ``designed'' according to the environment~\cite{Bianco2017}.

Similar $FOL\rightarrow UNF$ and $UNF\rightarrow AGG$ transitions are observed also in the proteins $A_1$, $A_2$, $D$, $E$ and $F$ (not shown here). It is interesting to observe that, although proteins $A_0$, $A_1$ and $A_2$ share the same native structure (the sequence of each proteins has been obteined with an independent design procedure), the concetration threshold fot the $FOL\rightarrow UNF$ and $UNF\rightarrow AGG$ transitions are different in each case.

\section{Water-mediated protein-protein interaction }

\begin{figure}
\centering
\includegraphics[width=0.7\textwidth]{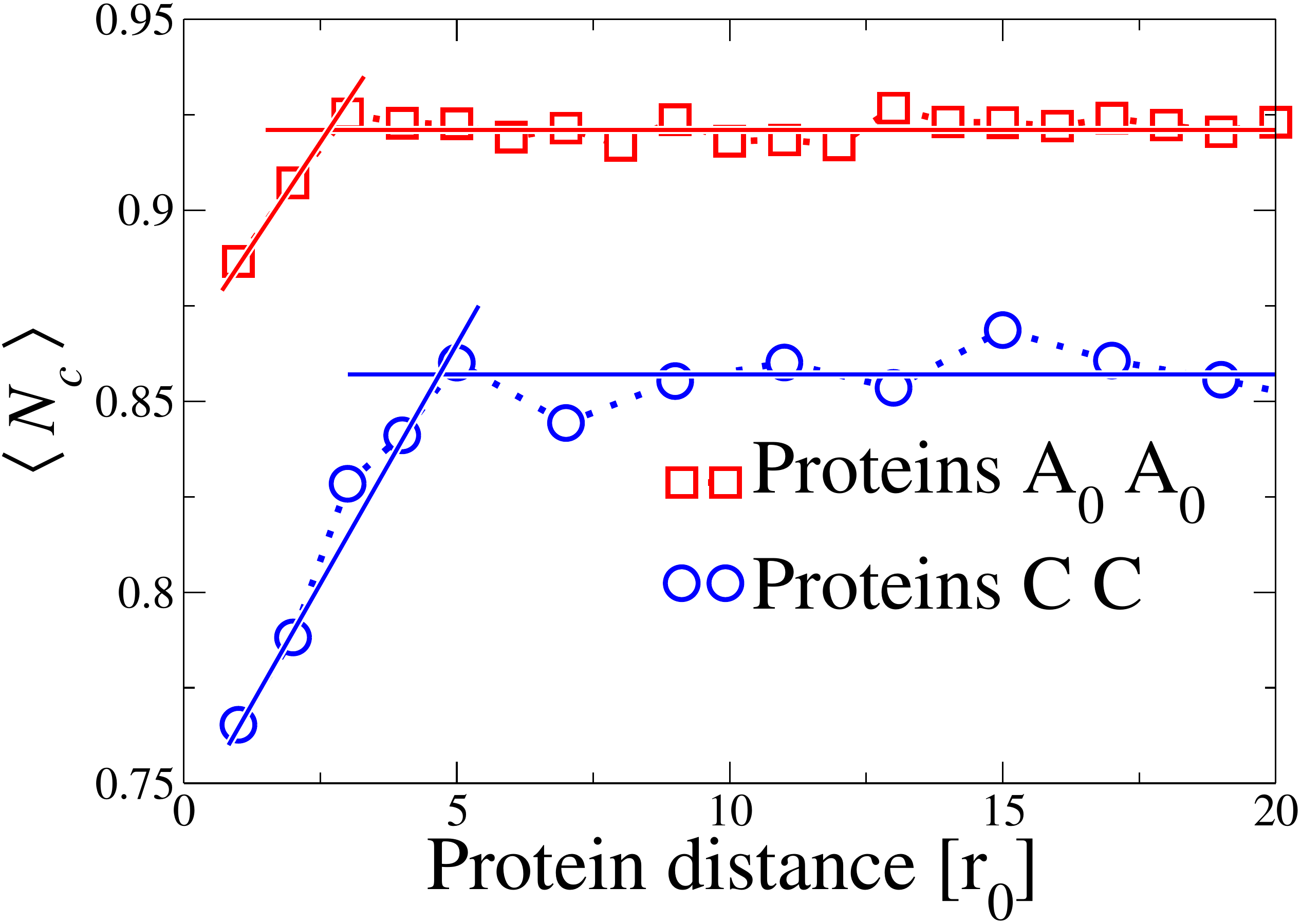}
\caption{Average number of native contacts $\langle N_c \rangle$ for the binary solutions with i) two proteins $A_0$ (red squares); ii) two proteins $C$ (blue circles). Data are plotted as function of the minimum distance between the two proteins $d_{pp}$. Lines are guides for the eye showing the increasing trend of $\langle N_c \rangle$ at smaller values of $d_{pp}$, and the constant value of $\langle N_c \rangle$ at larger $d_{pp}$. The intersection  between the lines identify the interaction radius of the proteins. The protein unfold at distance 2.5 for A and 5 for C both close to the average protein-proteins distances at the  $FOL\rightarrow UNF$ transition concentrations}
\label{fig:folding_vs_distance}
\end{figure}

\begin{figure}
\leftskip 1.7 cm 
\textbf{\large (a)\hspace{6.9 cm} (b)}\\
\vspace{- 0.5 cm}
\centering
\includegraphics[scale=0.7]{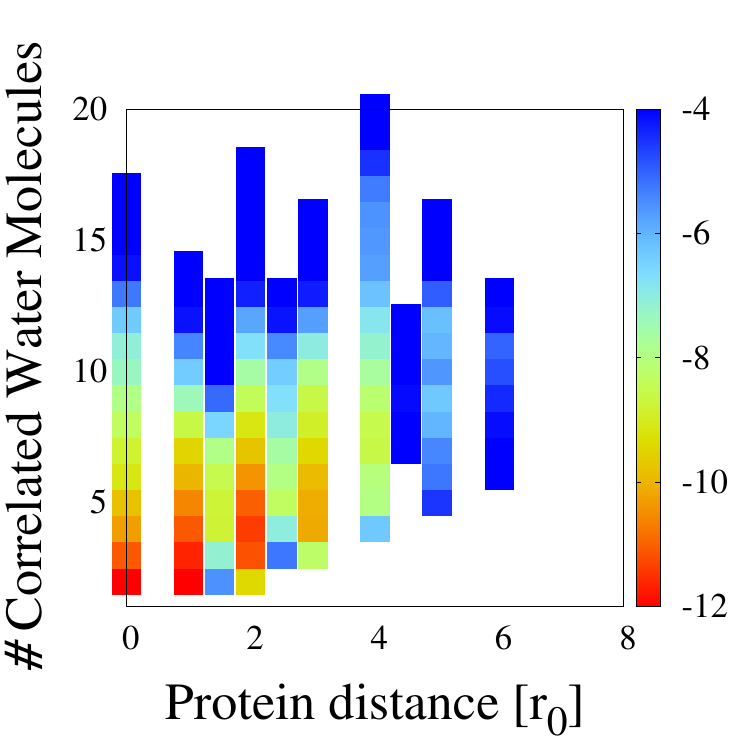}
\includegraphics[scale=0.7]{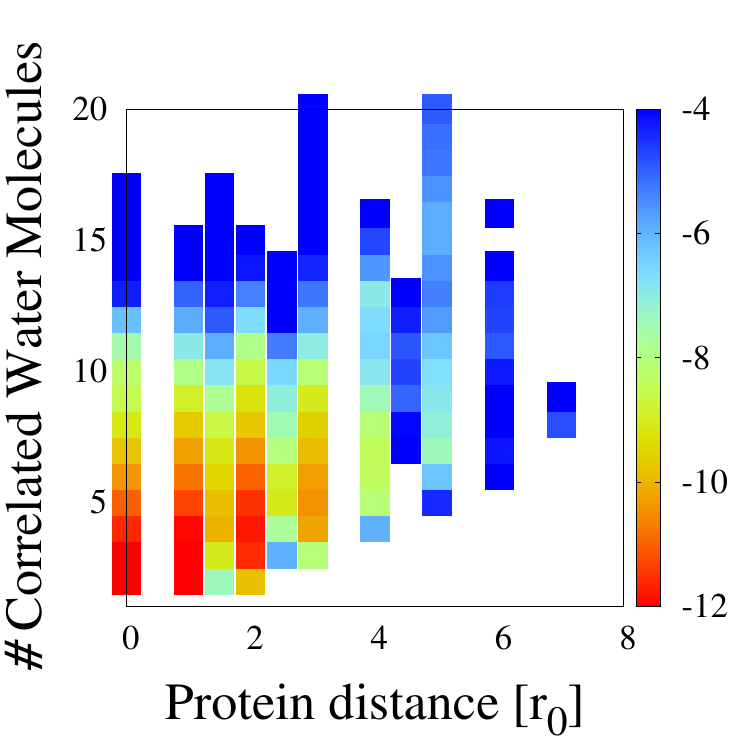}\\
\leftskip 1.7 cm 
\vspace{0.5 cm}
\textbf{\large (c)\hspace{6.9 cm} (d)}\\
\vspace{- 0.5 cm}
\centering 
\includegraphics[scale=0.7]{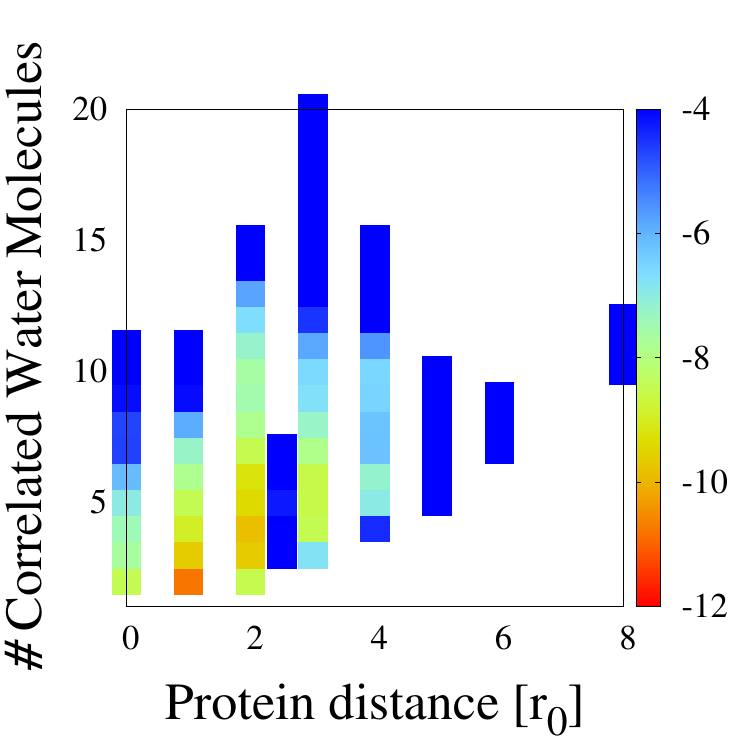}
\includegraphics[scale=0.7]{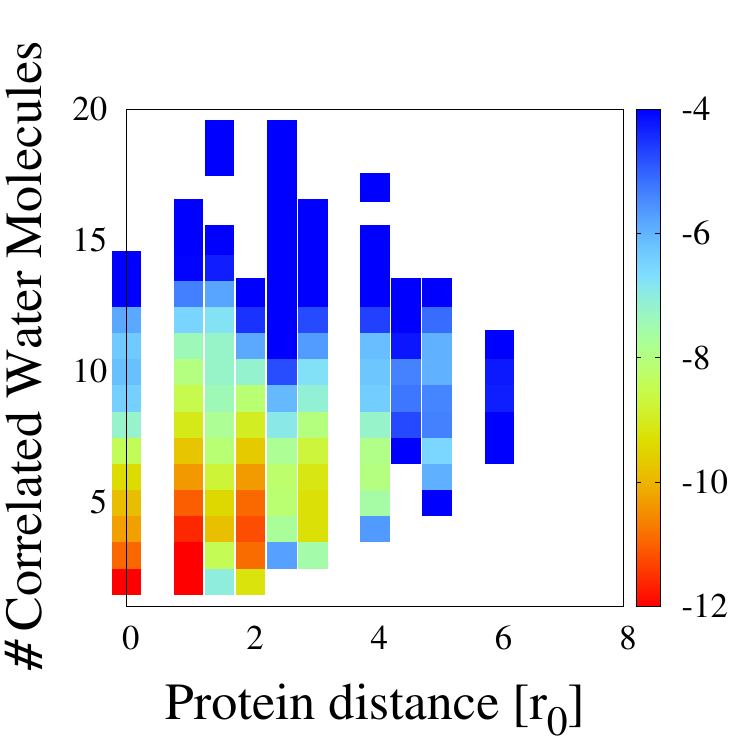}
\caption{Negative logarithm of the probability distribution of the clusters of statistically correlated water molecules in contact with two proteins, as function of the minimum protein distance and the number of water molecules belonging to the cluster. Following Ref. \cite{Bianco2019}, two neighbour bonding variables $\sigma_{ij}$ and $\sigma_{ji}$, such that $ \sigma_{ij} = \sigma_{ji}$,  belong to the same cluster with probability $p\equiv 1 - \exp (- \mathscr{J}/k_BT)$, where $\mathscr{J}$ is the specific interaction between $\sigma_{ij}$ and $\sigma_{ji}$. On average, we assume that an entire water molecules belong to a cluster any four bonding indices (since any water molecules is described by four bonding indices). 
(a) Clusters' distribution between proteins $A_0$ folded. (b) Clusters between proteins $A_0$ unfolded. (c) Clusters between proteins C folded. (d) Clusters between proteins C unfolded. The proteins at distance 2.5 for $A_0$ and 5 for C have clusters and that is the distance at which they unfold}
\label{fig:water_cluster}
\end{figure}

In this section we focus on the protein-protein interaction mediated by water molecules, for binary systems. In particular, we consider the cases $A_0$--$A_0$ proteins and $C_0$--$C_0$ proteins (homogeneous systems).
In Fig. \ref{fig:folding_vs_distance} we report the average number of native contact $\langle N_c \rangle$ \footnote{The average is calculated over all conformations hence the maximum average value will smaller compared to the global minimum that is $1$ for all proteins.} as function of the minimum protein distance \footnote{The minimum protein distance is the minimum value between all the possible distances among any amino acid of the first protein and any amino acid of the second protein.}. We observe that the value of $\langle N_c \rangle$ is constant for a wide range of of protein distances, with higher or lower values (respectively for the systems $A_0$--$A_0$ and $C$--$C$) reflecting the width of the free energy minima and hence the intrinsic stability of the native conformation. The interesting feature in Fig. S\ref{fig:folding_vs_distance} occurs when $\langle N_c \rangle$ starts decreasing linearly when the protein gets close to each other. These results demonstrate that the proteins start to unfold before interacting directly. Moreover, the transition distances correlate with the protein stability as $A_0$ (red square points), being overall more stable than the protein $C$ (blue circle points), show an interaction radius smaller ($\sim3r_0$) with respect the one of protein $C$ ($\sim5r_0$).
Our hypothesis is that the distance under which $\langle N_c \rangle$ decreases can be considered as the water-mediated the interaction radius of a protein. With this respect, following a recent percolation mapping \cite{Bianco2019},  we have performed a preliminary analysis of the extent of ``water statistical fluctuations" at the protein interface, depending on the protein folded/unfolded state and on the protein-protein distance. Such an extent is a measure of the correlation length in water and quantify the perturbation exerted by the protein on the surrounding water. Our data, shown in Fig. \ref{fig:water_cluster} the Supplementary Information, reveal an increase of the water fluctuations when two proteins unfold upon approaching each other. 
It is also important to notice that the transition distances are close to the distance between proteins at the  $FOL\rightarrow UNF$ transition concentrations.
Finally, the transition distances correlate with the protein stability as $A_0$ (red square points), being overall more stable than the protein $C$ (blue circle points), show a an interaction radius smaller ($\sim3r_0$) with respect the one of protein $C$ ($\sim5r_0$).

\section{Conclusions}

We have presented a computational study on the competition between folding and aggregation of proteins in homogeneous solutions. 
By means of an efficient coarse-grain model we have designed a series of proteins according to the water environment at ambient condition. Then, we have tested the capability of each designed protein to fold alone, and in presence of multiple copies (i.e. changing the protein concentration). 
The main conclusion of this work is that proteins tend to fold uninfluenced by the presence of other proteins in the solution provided that their concentration is below their specific unfolding concentration $c_{FOL \rightarrow UNF}$. Our simulations predict an unexpected and not previously observed role of the water in the inducing the unfolded regime $UNF$ that is a precursor of the fully aggregated state $AGG$. We believe that such prediction should be testable first in more detailed protein models and supports the need for new intriguing experiments. 

Correlated to our study, there is an extensive literature about the role of cellular crowding on aggregation and folding. A sample of pioneering works in the field are~\cite{VandenBerg1999,Gorensek-Benitez2017,Schummel2016,Feig2017,Gorensek-Benitez2017,Zhou2004a}. The central message of these studies is that the role of the steric crowding does not significantly affect the folding. However, when globular proteins replace crowding agents, the behaviour of the system becomes difficult to explain because of the influence of protein-protein. Our results offer a qualitative description of such an influence, separating the role of water, protein and steric interactions at different concentrations.

\subsection*{Acknowledgements}

V.B. acknowledges the support from the Austrian Science Fund (FWF) project  M 2150-N36 and from the European Commission through the Marie Skłodowska-Curie Fellowship No. 748170 ProFrost. 
V.B. and I.C.  acknowledge the support from the  FWF project P 26253-N27. Simulations have been performed using the Vienna Scientific Cluster (VSC-3).



\clearpage

\centering
{\Large Supplementary Information}

\clearpage

\begin{figure}
\leftskip 1 cm {\bf (a)\hspace{6.5 cm}(b)}\\
\centering
\includegraphics[scale=0.23]{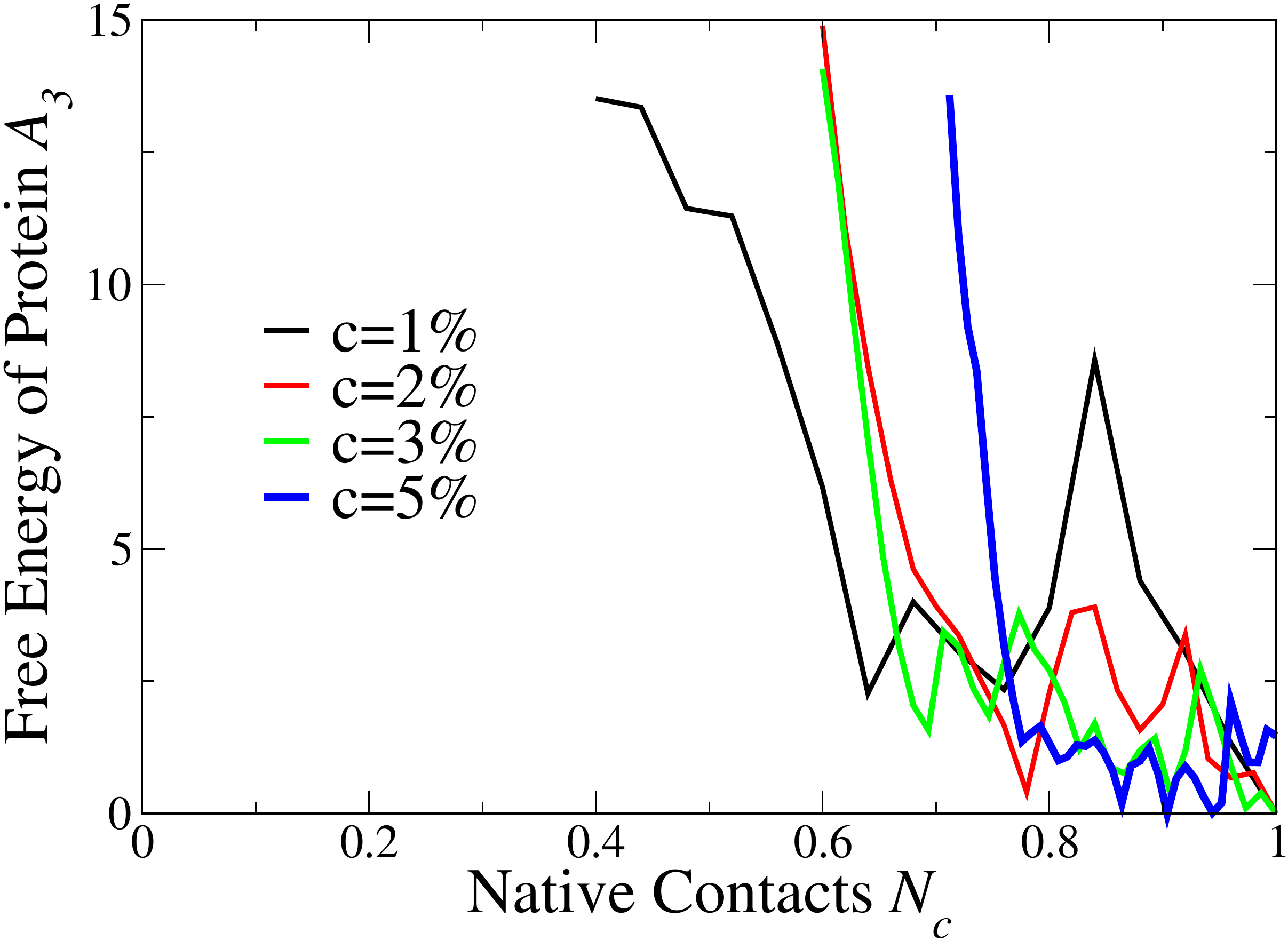}
\includegraphics[scale=0.23]{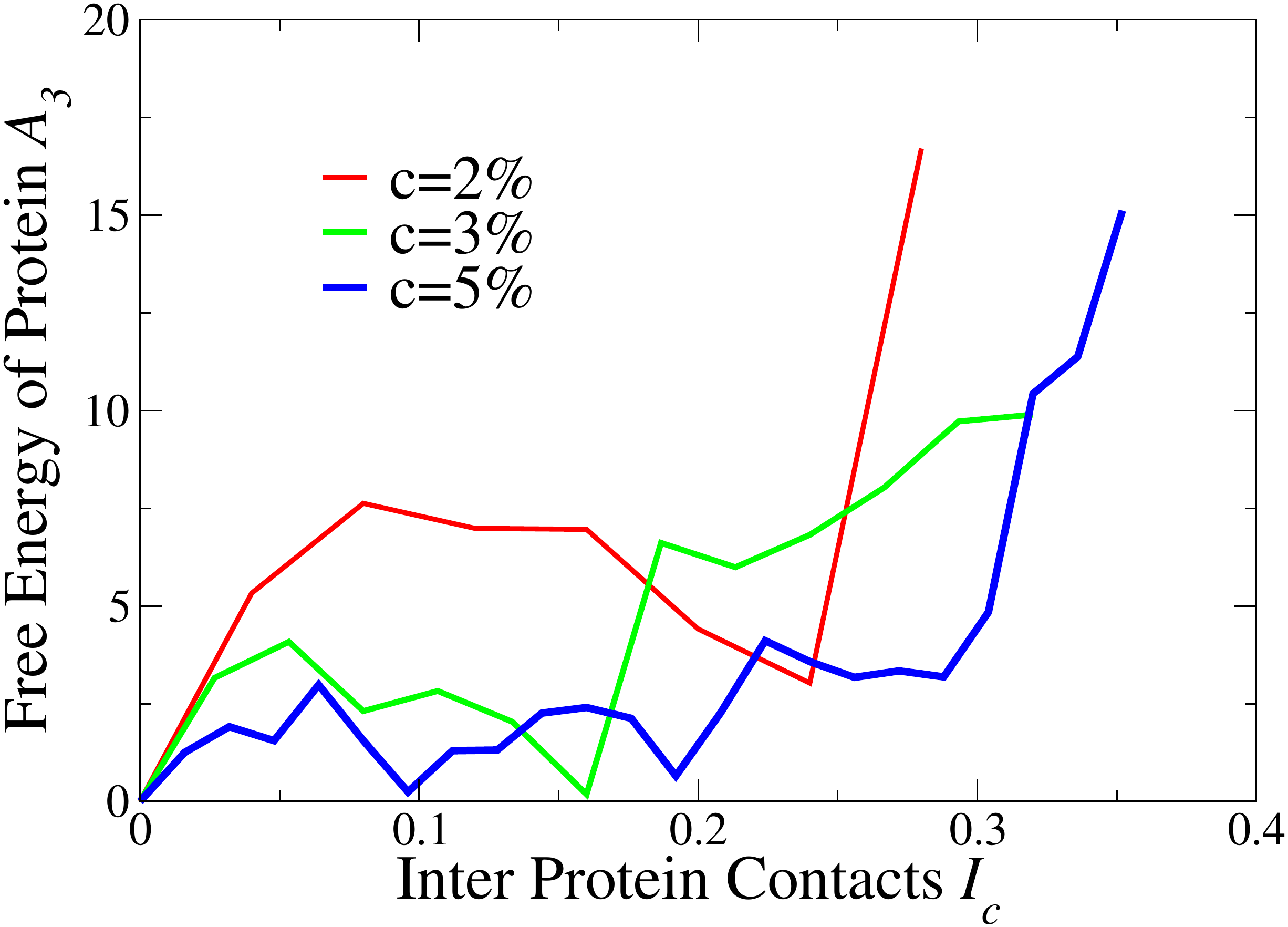}
\caption{Free energy profiles $F(N_c)$ (a) and $F(I_c)$ (b), function respectively of  $N_c$ and $I_c$ for the proteins $A_3$. We designed the sequence of protein $A_3$ switching off all the water-water interaction terms in the hydration shell. Protein $A_3$ is not surprisingly less stable than the sequence designed with explicit water \cite{Bianco2017}. The data show the disappearance of the $UNF$ state and the direct transition to the $AGG$ state. Moreover, the $FOL \rightarrow AGG$ transition takes place at much lower concentrations with respect to the case where the hydration water is explicitly accounted for (in the present case as low as 2\%). Hence, the hydration water acts as a barrier against the aggregation.}
\label{fig:implicit_solvent}
\end{figure}

\end{document}